\documentclass[epj]{svjour}
\usepackage{graphics}
\usepackage{amsmath}
\usepackage{amssymb}
\usepackage{amsfonts}
\usepackage{cuted}
\usepackage{lipsum} 
\usepackage{bibspacing}

\usepackage{url}
\usepackage[colorlinks, citecolor=blue,anchorcolor=blue,menucolor=blue, 
linkcolor=blue,filecolor=blue,runcolor=blue,urlcolor=blue,frenchlinks=true]{hyperref}

\begin{document}
\title{\centerline{The x-dependence of hadronic parton distributions:} 
\vskip 0.2cm
\centerline{A review on the progress of lattice QCD} \vskip 0.5cm}

\titlerunning{\,\,The x-dependence of hadronic parton distributions}
\authorrunning{Martha Constantinou\,}

\author{\centerline{ \bf Martha Constantinou\inst{} \thanks{marthac@temple.edu} }}
\institute{\centerline{  Temple University}}
\date{\hspace*{6.4cm} {October 5, 2020}\vspace{1cm}}
%
\abstract{
In this article, we review recent lattice calculations on the $x$-dependence of parton distributions, with the latter providing information on hadron structure. These calculations are based on matrix elements of boosted hadrons coupled to non-local operators and can be related to the standard, light-cone distribution functions via an appropriate factorization formalism. There is significant progress in several directions, including calculations of flavor singlet parton distribution functions (PDFs), first calculations of generalized parton distributions (GPDs), as well as the implementation of some of the approaches for the transverse-momentum-dependent PDFs (TMD PDFs). This new direction of lattice calculations is particularly interesting for phenomenological fits on experimental data sets, as the lattice results can help to improve the constraints on the distribution functions.
\PACS{{}{}
     } 
} 

\maketitle

\section{Introduction}
\label{intro}

The quest for an in-depth understanding of hadrons and their interactions dates back to the mid-20$^{\rm th}$ century. Pro-gress was rapidly made in both the experimental and theoretical frontiers, in particular after the formulation of QCD, that lead to the systematic and quantitative study of hadrons and their structure. 

The QCD factorization formalism developed for collider processes has been the foundation for understanding hadrons in terms of their partonic content~\cite{Collins:1981uw,Collins:1989gx}. Such a mapping of the hadron structure is achieved via a set of quantities, that is, the parton distribution functions (PDFs)~\cite{Collins:1989gx}, the generalized parton distributions (GPDs)~\cite{Ji:1996ek,Radyushkin:1996nd,Mueller:1998fv}, and the transverse-momentum-dependent distributions (TMD PDFs)~\cite{Collins:2003fm,Ji:2004wu}. These distributions are number densities and have a probabilistic interpretation within the parton model. However, they are not directly measurable in experiments. The mechanism that enables access to distribution functions, and therefore to the structure of hadrons, is the asymptotic freedom of QCD. By virtue of asymptotic freedom, the cross-section of high-energy processes can be factorized into a hard part and a soft part. For example, the cross sections for inclusive unpolarized 
Deep Inelastic Scattering (DIS) can be written as
\begin{eqnarray}
\label{eq:cross_section}
\sigma_{\rm DIS}(x,Q^2) 
&=& \sum_i \big[ H_{\rm DIS}^i \otimes f_i \big](x,Q^2)\,,\\[1ex]
\left[a \otimes b\right](x) &\equiv& \int_x^1 \frac{d\xi}{\xi}\, a\left(\frac{x}{\xi}\right)\, b(\xi) \,,
\end{eqnarray}
where $i$ represents all types of partons, that is quarks, anti-quarks and gluons. $x$ is the Bjorken scaling variable, and $Q^2$ represents the scale of the hard interaction. $H_{\rm DIS}$ is the hard part, it is process-dependent and calculable in perturbative QCD. $f_i$ is the non-perturbative part of the cross-section, characterizing the structure of hadrons. In the case of the unpolarized DIS, the relevant distribution function is the spin-averaged (or unpolarized) PDF for the $i^{\rm th}$ type of partons. Typically, both functions $H_{\rm DIS}$ and $f_i$ are given in the $\overline{\rm MS}$ scheme. Unlike $H$, the distribution functions are universal. Literally speaking, the distribution functions are not process-independent, as they depend on the renormalization scale, which varies in different processes. However, such a dependence does not pose difficulties, as the data sets can be evolved in the same scale using perturbation theory. Therefore, data sets from different processes can be analyzed together within the framework of global analysis. Factorization expressions similar to Eq.~(\ref{eq:cross_section}) exist for other high-energy processes, such as polarized DIS and Drell-Yan. Their corresponding $H$ function is known up to next-to-leading order (NLO), or next-to-next-to-leading order (NNLO), depending on the process.

The PDFs, GPDs, and TMD PDFs are light-cone correlation functions and cannot be accessed from the Euclidean formulation of lattice QCD. Partial information is obtained through the Mellin moments of distribution functions, and, in principle, one can reconstruct the parton distributions using an operator product expansion (OPE). Practically, a proper and exact reconstruction is an impossible task due to the computational challenges to calculate reliably high moments: the signal-to-noise rapidly decreases, and an unavoidable power-law mixing occurs beyond the third non-trivial moment. 

Alternative approaches to Mellin moments have been pursued, starting with a method based on the hadronic tensor~\cite{Liu:1993cv,Liu:1998um,Liu:1999ak} already in the 1990's. Other proposals include auxiliary quark field approaches~\cite{Detmold:2005gg,Braun:2007wv}, as well as methods to obtain high Mellin moments using smeared operators~\cite{Davoudi:2012ya}. The development of the quasi-PDFs approach using Large-Momentum Effective Theory (LaMET) has been a pivot point for calculating $x$-dependent quantities from lattice QCD~\cite{Ji:2013dva,Ji:2014gla,Ji:2020ect}. This approach created an intense enthusiasm in both the phenomenological and lattice communities, which led to other proposals, such as the current-current correlators approach~\cite{Ma:2014jla,Ma:2014jga,Ma:2017pxb}, the pseudo-PDFs~\cite{Radyushkin:2016hsy}, and a method based on OPE~\cite{Chambers:2017dov}.

This article reviews lattice calculations on extracting the $x$-dependence of distribution functions utilizing some of the aforementioned approaches, for which there has been recent progress. Some of the calculations prior to 2019 can be found in the proceedings of the Lattice Symposium 2018~\cite{Monahan:2018euv}. An extensive review of the first calculations up to 2018 can be found in Ref.~\cite{Cichy:2018mum}. 
The main focus of this review is the presentation of physical results that are also interesting to the wider theoretical community. Therefore, the technical details and heavy equations are avoided, in order to increase readability. However, selected important technical aspects that affect the reliability of lattice results are critically discussed. We present calculations that appeared in the literature up to October 1, 2020.

The main part of this article is organized in three sections. Section~\ref{sec:PDFs} presents calculations on PDFs, Section~\ref{sec:GPDs} on GPDs, and last but not least, Section~\ref{sec:TMDs} focuses on TMD PDFs. Finally, we provide closing remarks in Section~\ref{sec:Discussion}.

\vspace*{1.5cm}
\section{PDFs}
\label{sec:PDFs}

Most of the information on hadron structure comes from the intense effort to determine PDFs with quantifiable and reliable uncertainties. Among the distribution functions, PDFs are the easiest to obtain, as they are one-dimensional objects that depend only on the momentum fraction carried by the partons. 

The PDFs are classified as unpolarized (spin-averaged), helicity (spin-dependent) and transversity (transverse-spin-dependent), and each can be obtained through the factorization of the cross-section from different high-energy processes. However, there is still missing information, either because it is challenging to extract certain PDFs, or because some kinematic regions are not easily accessible. Therefore, input from lattice QCD is crucial, and lattice results can either serve as predictions, or as additional constraints on the phenomenological analyses. In this section, we will discuss the current landscape of lattice results, highlighting their $x$-dependence in the presentation, rather technical details.

\subsection{Leading-twist PDFs}
\label{ssec:twist2PDFs}

The non-perturbative part of the cross-section may be expanded in terms of the large energy scale of the process under study, $Q$. Taking as an example the unpolarized PDFs in Eq.~(\ref{eq:cross_section}), one can write the expansion in twist (mass dimension minus spin)
\begin{equation}
f_i = f_i^{(0)} + \frac{f_i^{(1)}}{Q} + \frac{f_i^{(2)}}{Q^2} \cdots \,,  
\label{eq:twist}
\end{equation}
where $f_i^{(0)}$ is the leading-twist (twist-2) contribution, while $f_i^{(1)}$ and $f_i^{(2)}$ are the twist-3 and twist-4 contributions, respectively. The leading-twist contributions have probabilistic interpretation, while the higher-twist are sensitive to the soft dynamics. In most of the studies, the main focus is on leading twist contributions, which are the easiest to isolate. While the higher-twist corrections are less-studied, there is a growing interest including lattice calculations (See, also, Sec.~\ref{ssec:twist3}).

Progress has been made in several directions, and some will not be covered in this article due to space limitations. This includes, but is not limited to, the kaon distribution amplitude (DA)~\cite{Lin:2020ssv}, the feasibility of extracting the $x$-dendendence of PDFs for the $\Delta^+$ baryon~\cite{Chai:2020nxw}, and exploration of machine-learning methods for the pion and kaon DAs~\cite{Zhang:2020gaj}. We also refer the Reader to a series of papers on moments of distribution amplitudes for the kaon~\cite{Bali:2019dqc}, the lowest-lying baryon octet~\cite{Bali:2019ecy}, and the double parton distributions of the pion~\cite{Bali:2020mij}. Preliminary results on the second moment of the pion DA using heavy quark OPE, can be found in the proceedings of Ref.\cite{Detmold:2020lev}.

The main effort of lattice calculations up to 2018 was to obtain $x$-dependent PDFs directly at the physical point. Once that was achieved, a shift towards understanding systematic uncertainties, as well as, using the methodology to extact more complicated quantities has been observed. This change of what is considered a priority in the field, is reflected in this review.

\vspace*{0.5cm}
\subsubsection{quasi-PDFs framework}
\label{sssec:qPDFs}

The rapid progress in the field of $x$-dependent distributions from lattice QCD, became possible with the pioneering work of X. Ji in 2013~\cite{Ji:2013dva} on a new approach to extract PDFs from a Euclidean lattice. While this is not the first approach to be considered, it clearly renewed the interest in that direction, and marked the beginning for a new era in lattice calculations. This approach, the so-called quasi-distributions approach, demonstrated that matrix elements of fast-moving hadrons coupled with bilinear non-local operators can be related to light-cone distributions, via a multi-step process. As an example, we write the matrix element in the forward limit which leads to PDFs, that is
\begin{equation}
{\cal M}(P_3,z,\mu) = Z_\Gamma(z,\mu) \langle H(P_3) | \bar{\psi}(z)\Gamma W(z) \psi(0) |H(P_3) \rangle\,,
\label{eq:ME}
\end{equation}
where the hadron is boosted with momentum $P_3$ chosen to be in the $z$ direction. The operator contains a Wilson line of length $z$ parallel to $P_3$, and has Dirac structure $\Gamma=\gamma^0,\,\gamma^3\gamma^5,\,\sigma^{3j}$ ($j=1,2$), for the unpolarized~\footnote{In the first studies, the operator $\gamma^3$ was used for the unpolarized PDFs. However it was found that it mixes with the twist-3 scalar operator~\cite{Constantinou:2017sej} and abandoned.}, helicity and transversity PDFs, respectively. $Z_\Gamma(z,\mu)$ is the renormalization function defined at renormalization scale $\mu$. $Z_\Gamma(z,\mu)$ is calculated at each $z$ separately, and is scheme- and scale-dependence (except for the vector and axial currents at $z=0$). The quasi-PDFs are defined as the Fourier transform (FT) of ${\cal M}(P_3,z,\mu)$
\begin{equation}
\label{eq:qPDF}
\tilde{q} (x, P_3,\mu) = 2 P_3 \int_{-\infty}^\infty \frac{dz}{4\pi} e^{-i z x P_3} {\cal M}(P_3,z,\mu)\,.
\end{equation} 
The approach relies on the fact that the difference between quasi-PDFs and light-cone PDFs is nonzero only in the ultra-violet (UV) region, and can be calculated in continuum perturbation theory. For large, but finite $P_3$, the quasi-PDFs can be matched to their light-cone counterparts using a kernel calculated perturbatively in Large Momentum Effective Theory (LaMET)~\cite{Ji:2014gla}, that is
\begin{equation}
{\widetilde q}_\Gamma(x,P_3,\mu) {=} \hspace*{-0.1cm} \int_{-1}^{1} \frac{d\xi}{|\xi|}\,
    C_\Gamma\Big( \frac{x}{\xi}, \frac{\mu}{\xi P_3} \Big)\,
    q(\xi,\mu)
 {+} \mathcal{O}\bigg(\frac{m^2}{P_3^2},\frac{\Lambda_{\mbox{\tiny QCD}}^2}{x^2P_3^2}\bigg).
\end{equation}

Following the first calculations exploring the feasibility of the quasi-PDFs framework~\cite{Lin:2014zya,Alexandrou:2014pna,Alexandrou:2015rja,Chen:2016utp,Alexandrou:2016jqi}, $x$-dependent PDFs calculated directly at the physical point have been obtained~\cite{Alexandrou:2018pbm,Alexandrou:2018eet,Lin:2018pvv}. Several factors contributed to this success, such as, the use of momentum smearing~\cite{Bali:2016lva}, which offers better overlaps with the ground state. Another important development was the proposal~\footnote{The renormalization and mixing pattern were first presented to the community at the 2017 APS Topical Group on Hadronic Physics (GHP) Meeting~\cite{GHP}.} for a complete renormalization scheme~\cite{Constantinou:2017sej,Alexandrou:2017huk}, which was complemented by the proof of renormalizability of non-local operators to all orders in perturbation theory~\cite{Ji:2015jwa,Ishikawa:2017faj,Ji:2017oey,Wang:2017eel,Zhang:2018diq,Li:2018tpe}. Last but not least, a well-defined matching prescription appropriate for the various renormalization schemes, has been developed~\cite{Wang:2017qyg,Stewart:2017tvs,Izubuchi:2018srq,Liu:2018uuj,Liu:2018hxv,Alexandrou:2019lfo}. All this progress and the ability to calculate PDFs at the physical point, led to the need to refine the lattice estimates, with in-depth studies of various sources of systematic uncertainties.

\begin{figure}[h!]
\begin{center}
\resizebox{0.45\textwidth}{!}{\includegraphics{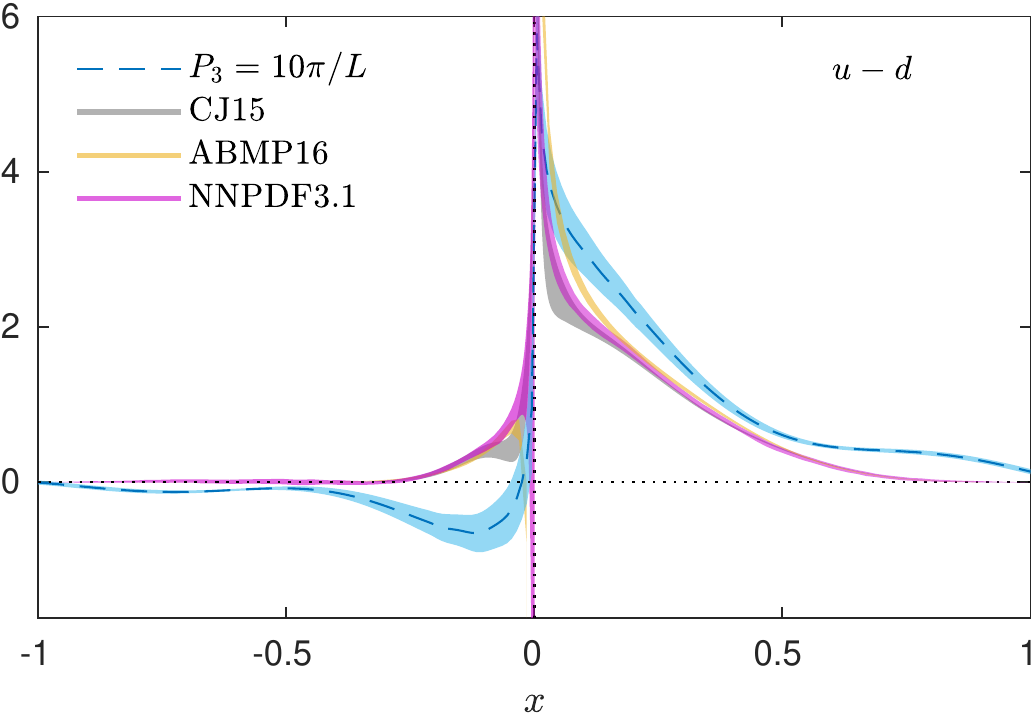}}
\resizebox{0.45\textwidth}{!}{\includegraphics{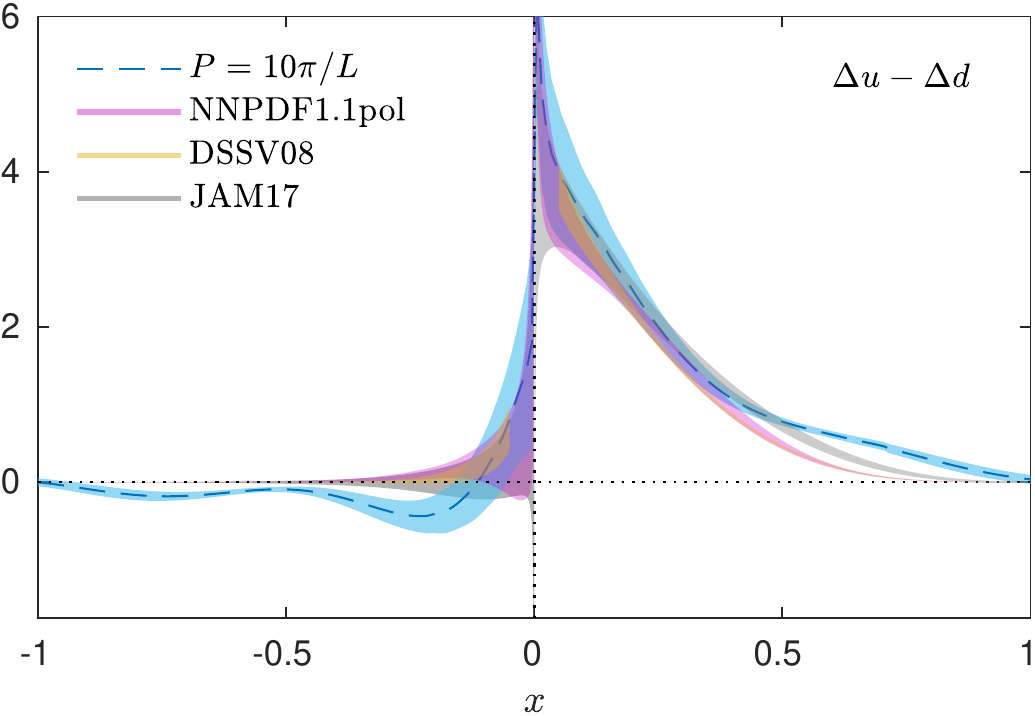}}
\resizebox{0.45\textwidth}{!}{\includegraphics{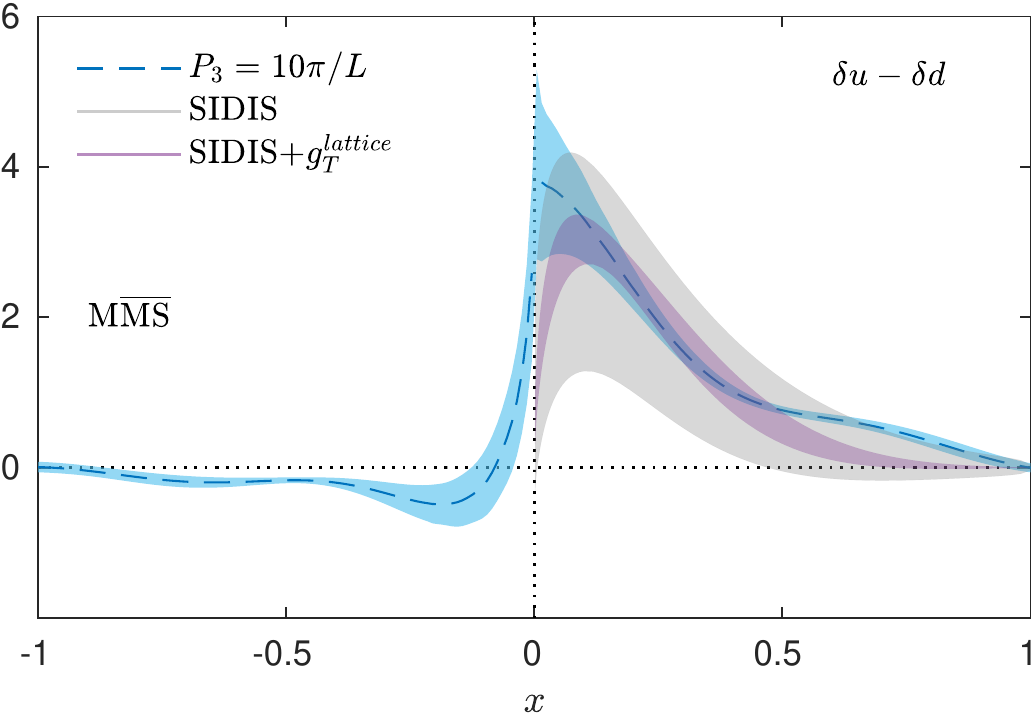}}
\caption{The unpolarized (top), helicity (center) and transversity (bottom) PDFs for the nucleon at the physical point and $P_3{=}1.38$ GeV. (blue curve). The global fits of Refs.~\cite{Alekhin:2017kpj,Ball:2017nwa,Accardi:2016qay} (unpolarized), Refs.~\cite{deFlorian:2009vb,Nocera:2014gqa,Ethier:2017zbq} (helicity), Refs.~\cite{Lin:2017stx} (transversity) are shown for qualitative comparison. Source: Ref.~\cite{Alexandrou:2019lfo}. Article published under the terms of the Creative Commons Attribution 4.0 International license.}
\label{fig:PDFpheno}
\end{center}
\end{figure}

A thorough investigation of systematic uncertainties in the unpolarized, helicity and transversity PDFs is presented in Ref.~\cite{Alexandrou:2019lfo}. The focus is on selected sources of systematic effects, as only one ensemble is used. The analysis is done using an $N_f=2$ twisted-mass fermions ensemble with physical pion mass and spatial extent of 4.5 fm ($48^3\times96$). Using four values of the source-sink separation, $T_{\rm sink}=0.75 - 1.13$ fm, excited-states effects are addressed with single-state fits, two-state fits, and the summation method. The conclusion from this analysis is that lattice data at $T_{\rm sink}$ below 1 fm are not sufficient to suppress excited-states, particularly as $P_3$ increases. Another effect that requires attention for matrix elements of non-local operators is finite-volume effects~\cite{Briceno:2018lfj}. Direct study of volume effects is not possible with one ensemble, but one can examine whether the renormalization functions, $Z$, exhibit volume effects, if obtained using multiple ensembles. In the work of Ref.~\cite{Alexandrou:2019lfo} the estimates of $Z$ are obtained using ensembles with different volume, that is 2.25 fm, 4.5 fm, and 6 fm. It is found that volume effects are small, and do not affect the final estimates of the PDFs. However, a proper study of volume effects on the matrix elements should be done, ideally at the physical point. Two other sources of uncertainties are presented: the reconstruction of the $x$-dependence via a FT, and the matching procedure. For the former, the standard discretized FT and the ``derivative method''~\cite{Lin:2017ani} are tested. While both methods suffer from the ill-defined inverse problem, the derivative method leads to uncontrolled uncertainties in the small-$x$ region (see discussion in Sec.~\ref{sssec:LCSs}). Regarding the truncation of the FT with respect to $z$, an optimal value of $z_{\rm max}$ requires that both the real and imaginary parts of the matrix elements are zero. In practice, this is not always observed, and the behavior of the matrix element in the large-$z$ is operator-dependent. The final choices for $z_{\rm max}$ in Ref.~\cite{Alexandrou:2019lfo} are 0.94 fm for the unpolarized, and 1.13 fm for the helicity and transversity. Different prescriptions for the matching are employed, with the quasi-PDFs defined in the $\overline{\rm MS}$, RI, ${\rm M}\overline{\rm MS}$ and ratio scheme. Tension is observed between some of the methods in the small- and large-$x$ regions, as expected. The final PDFs using $P_3=1.38$ GeV are shown in Fig.~\ref{fig:PDFpheno}. 
Phenomenological estimates are also shown for qualitative comparison. Focusing on the positive $x$-region, the lattice unpolarized PDF lies above the global fits, while there is agreement in the helicity PDF for $x\lesssim0.4-0.5$. The mild rise in the large $x$-region, could be due to the limitations of the FT, and advanced reconstruction methods must be investigated. Finally, the transversity PDF for $x\lesssim0.4-0.5$ is in agreement with the SIDIS results with or without the tensor charge constraint. The rise in the large-$x$ region is observed on all PDFs. The analysis of Ref.~\cite{Alexandrou:2019lfo} provide insight on selected sources of systematic uncertainties using simulations at the physical point. The bands in Fig.~\ref{fig:PDFpheno} include only statistical uncertainties, as the analysis confirms that the effects are either negligible, e.g., in $Z$, or require further investigations, such as, advanced reconstruction techniques.

\begin{figure}[h!]
\begin{center}
\resizebox{0.45\textwidth}{!}{\includegraphics{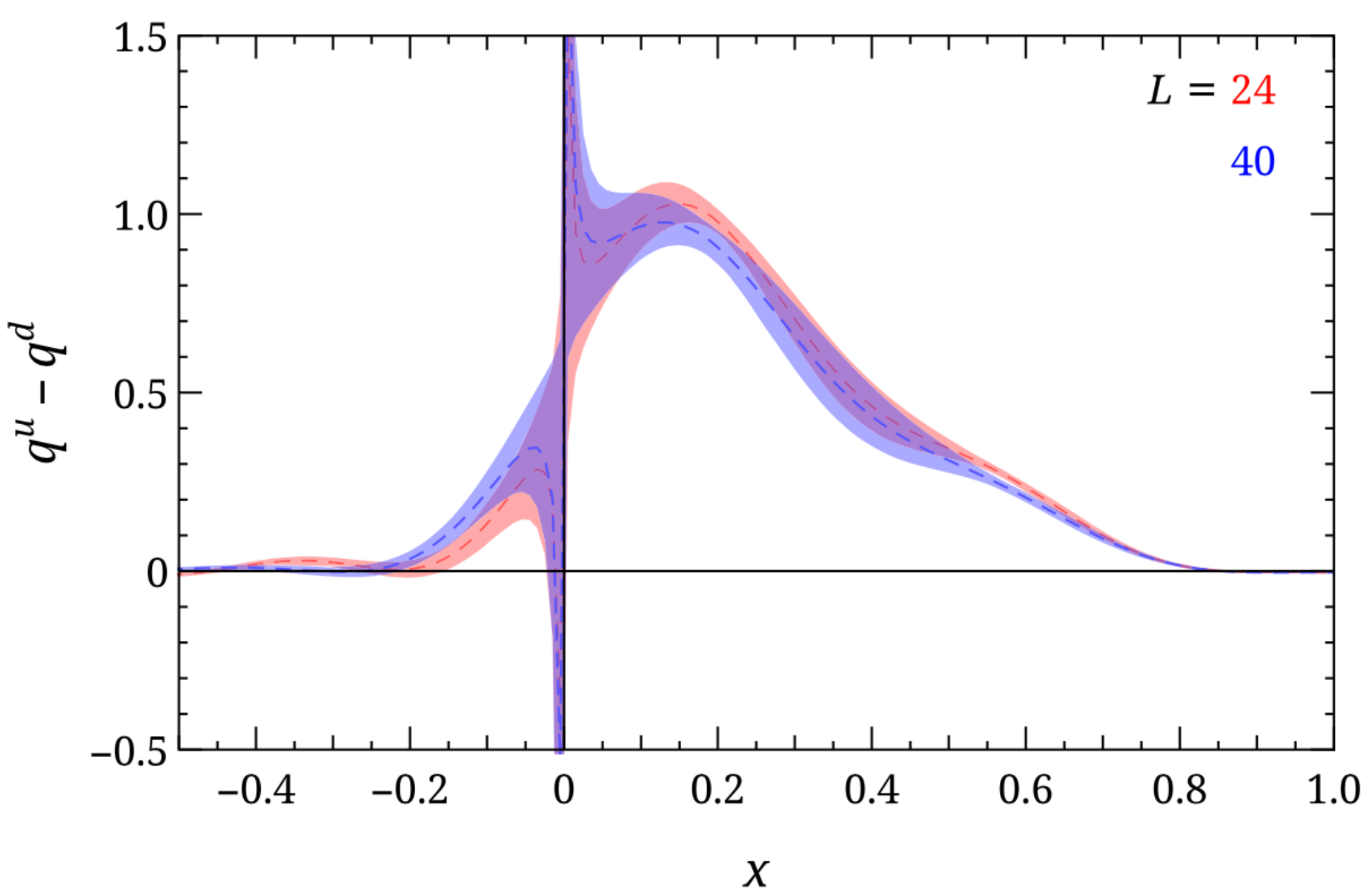}}
\resizebox{0.45\textwidth}{!}{\includegraphics{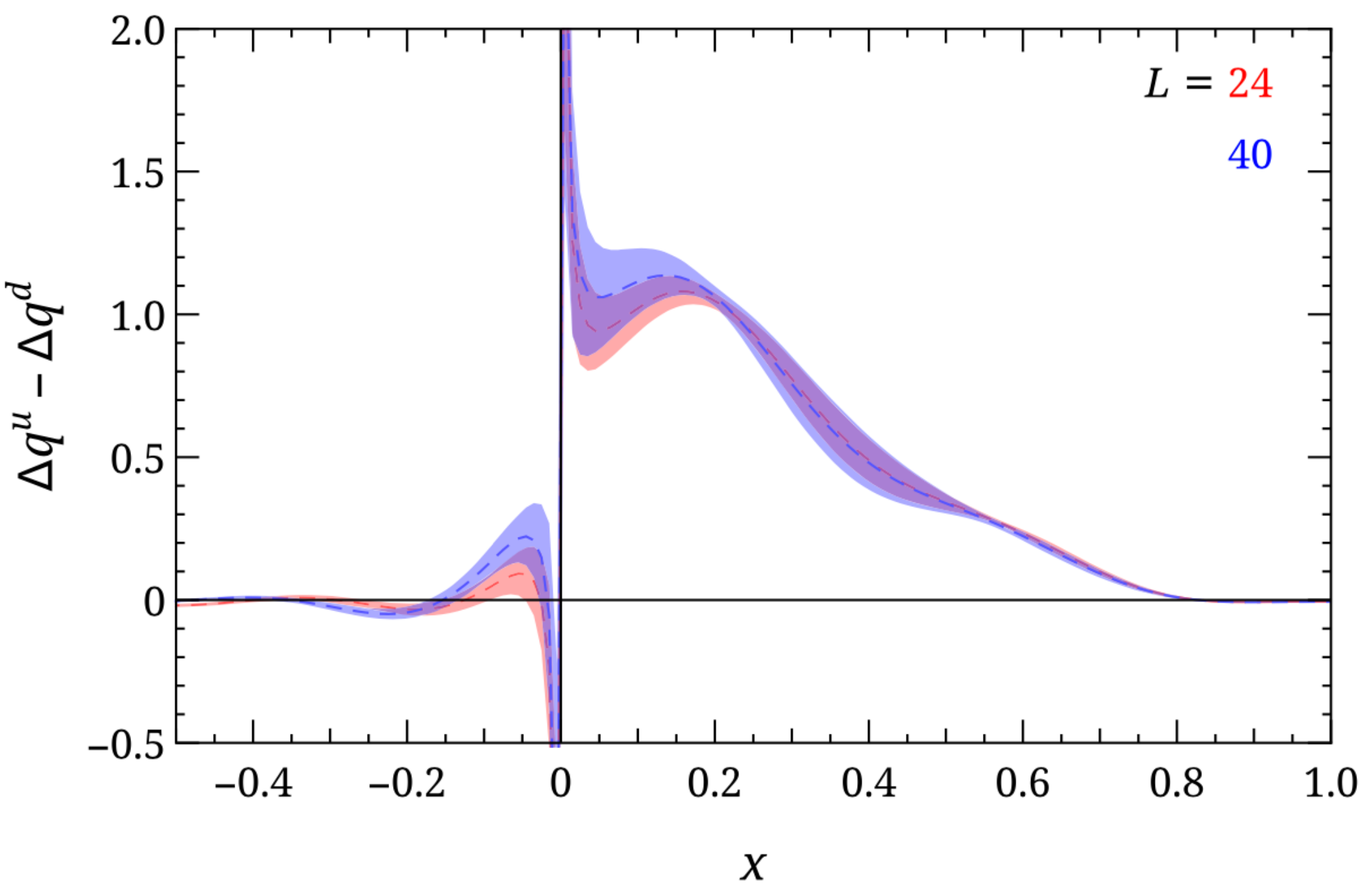}}
\caption{Unpolarized (top) and helicity (bottom) PDFs for the $L=2.88$ fm (red) and $L=4.8$ fm ensemble. Source: Ref.~\cite{Lin:2019ocg}. Reprinted with permission by the American Physical Society and SciPri.}
\label{fig:2019ocg}
\end{center}
\end{figure}

Finite-volume effects have been studied for the unpolarized and helicity PDFs in Ref.~\cite{Lin:2019ocg}. The calculation uses a mixed-action setup on three $N_f=2+1+1$ HISQ ensembles, with clover valence fermions. The pion mass of the ensembles is 220 MeV, and the lattices have spatial extent $L=2.88$, 3.84, and 4.8 fm. The momentum boost used is 1.3 for all three ensembles, and 2.6 GeV for the smallest- and largest-volume ensemble. Excited-states effects are quantified using four $T_{\rm sink}$ values between 0.72 - 1.08 fm, and the final results are presented using two-state fit. The matrix elements at the same $P_3$ from different ensembles, are found to be compatible, indicating that volume-effects are not significant for $m_{\rm val} L \in [3.3 - 5.5]$. Fig.~\ref{fig:2019ocg} shows the unpolarized and helicity PDFs using the $L=2.88$ fm and $L=4.8$ fm. For both cases, the curves are compatible for all regions of $x$. This is in agreement with the findings of Ref.~\cite{Alexandrou:2019lfo}, where the volume effects were tested on the renormalization functions, and found small.

Ref.~\cite{Fan:2020nzz} presents an analysis of systematic uncertainties also on a mixed action setup of clover fermions in the valence sector on $N_f=2+1+1$ HISQ fermions with pion mass 310 MeV. The lattice spacing is 0.042 fm and $L=2.7$ fm. While the physical volume of the ensemble is small, it is expected to have reduced discretization effects. As discussed above, there are indications that volume effects are not a major source of systematic uncertainties, at least for simulations at heavier pion mass. The 2pt-functions are calculated at 7 values of $P_3$ between 0 and 2.77 GeV, to understand the effect of excited-states contamination, using up to three-state fits. It is found that as $P_3$ increases, the fits become less reliable and the use of priors is needed to increase stability. For the 3pt-point functions, only $P_3=1.84,\,2.31$ GeV are used and three values of $T_{\rm sink}= 0.67 - 0.84$ fm. A 3-, 4-, 5-parameter two-state fit is applied, with the difference in the energy between the ground-state and first-excited state determined independently from the 2pt-functions fits. An agreement between the three fits is found, however the matrix elements from the 5-parameter fit are a factor of two more noisy.
\begin{figure}[h!]
\begin{center}
\resizebox{0.45\textwidth}{!}{\includegraphics{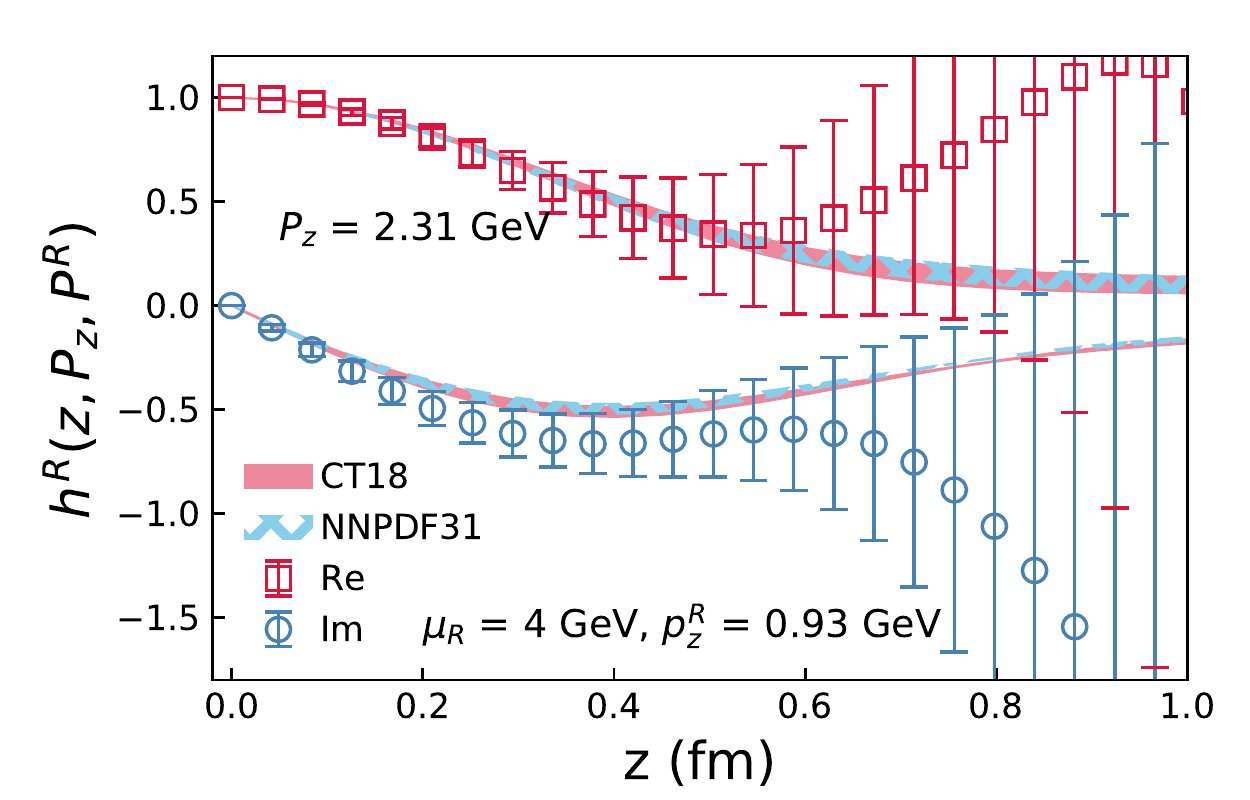}}
\resizebox{0.45\textwidth}{!}{\includegraphics{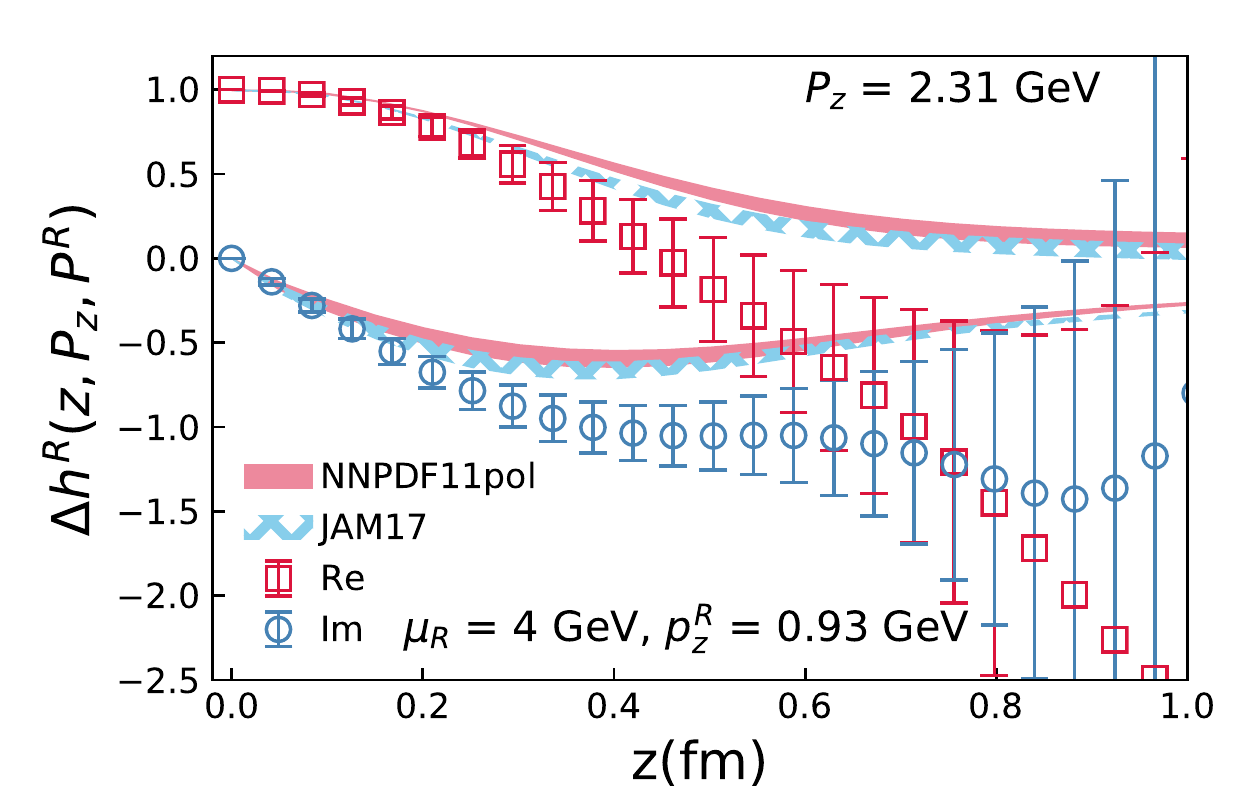}}
\caption{The renormalized matrix elements for the unpolarized (top) and helicity (bottom)  case for  $P_3=2.31$ GeV. The real and imaginary parts are shown with red and blue points, respectively. The global fits of NNPDF31~\cite{Ball:2017nwa} and CT18~\cite{Hou:2019efy} are shown for the unpolarized, and NNNPDF1.1pol~\cite{Nocera:2014gqa} and JAM17~\cite{Ethier:2017zbq} for the helicity. Source: Ref.~\cite{Fan:2020nzz}. Reprinted based on the arXiv distribution license.}
\label{fig:comp2lat_z}
\end{center}
\end{figure}
The renormalized matrix elements are compared in the coordinate space with the estimates from global analyses, as shown in Fig.~\ref{fig:comp2lat_z}. The real part of the unpolarized case shows agreement for all $z$ values shown, however, the errors grow above 50$\%$ at around $z\sim0.6$ fm, making the comparison beyond that point less reliable. The imaginary part has agreement up to $z\sim0.2$ fm, with large uncertainties for large $z$ values. The real part of the helicity matrix element is in agreement with global fits up to $z\sim0.3$ fm, while the imaginary part up to $z\sim 0.2$ fm. The large uncertainties and the long tails in the lattice data for $z>0.6$ fm, reveal the challenges of the inverse problems, as will be discussed below. Implementing a ratio renormalization prescription (pseudo-ITDs), the first moments of the PDFs are determined. More details can be found in Ref.~\cite{Fan:2020nzz}.

The pion valence PDF of Ref.~\cite{Izubuchi:2019lyk} follows the quasi-PDFs approach using a mixed action of clover fermions in the valence sector and $N_f=2+1$ HISQ fermions with pion mass 300 MeV. The volume is $48^3\times 64$, corresponding to a spatial extent of $2.9$ fm ($a=0.06$ fm). Three values of $P_3$ are used, namely, 0.86, 1.29, and 1.72 GeV, and there is agreement between $P_3=1.72$ GeV and the phenomenological fits of the JAM Collaboration~\cite{Barry:2018ort}. Results from Ref.~\cite{Izubuchi:2019lyk} are included in Fig.~\ref{fig:pion_QPFG_pPDF}. 
\begin{figure}[h!]
\begin{center}
\resizebox{0.45\textwidth}{!}{\includegraphics{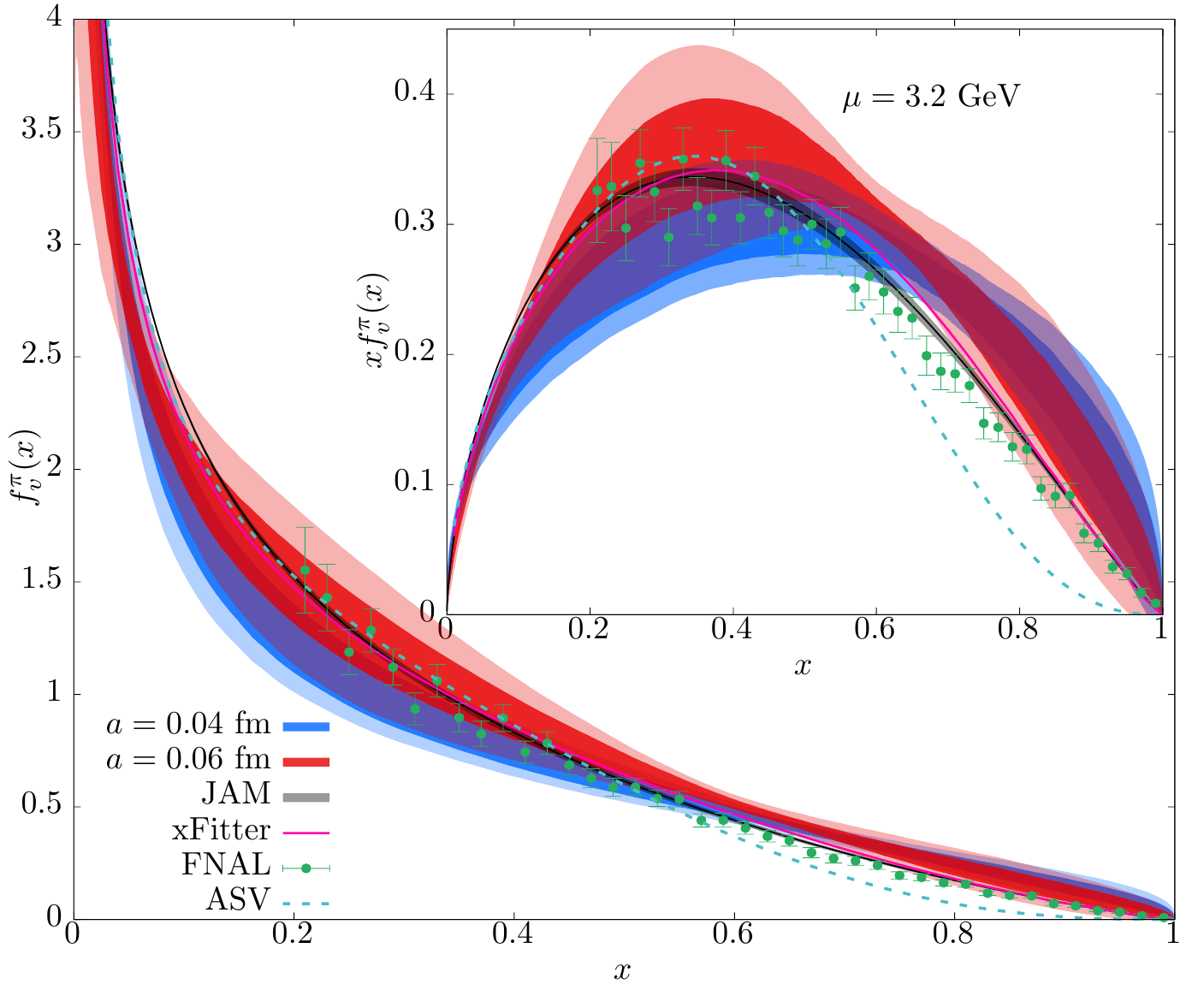}}
\caption{Pion valence PDF extracted using ensembles with $a=0.06$ fm (red) and $a=0.04$ fm (blue). The statistical (systematic) error is shown with dark (light) color band. The FNAL E-0615~\cite{Conway:1989fs} (green points), ASV~\cite{Aicher:2010cb} (green dashed line), JAM~\cite{Barry:2018ort} (black band) and xFitter analysis~\cite{Novikov:2020snp} (purple line) are also shown.  Insets: $x f_v^\pi(x)$. Source: Ref.~\cite{Gao:2020ito}.  Reprinted based on the arXiv distribution license.}
\label{finalpdf}
\end{center}
\end{figure}
The calculation is extended in Ref.~\cite{Gao:2020ito}, in which the statistics is increased, and another ensemble with $a=0.04$ fm and spatial extent 2.6 fm is added. The pseudo-ITD approach (see Sec.~\ref{sssec:pPDF}) is also explored. To this end, several values of $P_3$ are used, up to 2.15 GeV for the $a=0.06$ fm ensemble, and 2.42 GeV for the $0.04$ fm ensemble. Systematic uncertainties related to excited-states contamination using $T{\rm sink}$ up to 0.72 fm, have been studied. Other systematic uncertainties are investigated related to renormalization, higher-twist effects for low momenta, and study of the functional form in the large-$x$ region, using data with $z$ up to 1 fm. The large-$x$ region is of particular interest, as models predict different behavior of the form $(1-x)$, or $(1-x)^2$. The valence pion PDF is shown in Fig.~\ref{finalpdf} for the two ensembles. The comparison of the lattice data with other predictions and experimental data shows that there is agreement in the intermediate-$x$ region. For $x>0.7$, the $a=0.06$ fm data are in agreement with the FNAL estimates~\cite{Conway:1989fs}, when systematic uncertainties are included to the lattice data. Also, it is found that fits of the form $(1-x)^\beta$ applied to the lattice data finds $\beta$ anywhere between [1 - 2], and therefore, there is no clear preference in one of the two functional forms proposed. Comparison with other lattice calculations, reveals that the large-$x$ as obtained from Ref.~\cite{Gao:2020ito} is in agreement with Ref.~\cite{Izubuchi:2019lyk}, but disagrees with the results of Refs.~\cite{Joo:2019bzr,Sufian:2019bol} (see Fig.~\ref{fig:pion_QPFG_pPDF}).

Other calculations on quasi-PDFs and comparison with alternative approaches are presented in Sec.~\ref{sssec:pPDF} for the nucleon and Sec.~\ref{sssec:LCSs} for the pion.

\subsubsection{pseudo-ITD framework}
\label{sssec:pPDF}

An alternative framework to quasi-PDFs is the pseudo-ITDs~\footnote{The term pseudo-PDFs is also used.} developed by A. Radyushkin~\cite{Radyushkin:2016hsy,Radyushkin:2017cyf,Radyushkin:2017lvu,Radyushkin:2017sfi,Radyushkin:2018cvn,Radyushkin:2018nbf,Radyushkin:2019owq,Radyushkin:2019mye}. The common feature with quasi-PDFs is the use of the same matrix elements $\mathcal{M}$ of boosted hadrons coupled to non-local operators (Eq.~(\ref{eq:ME})). In the pseudo-ITDs approach, the analysis uses the Lorentz invariants, $z^2$ and $\nu\equiv -p\cdot z$, instead of $z$ and $P_3$. The variable $\nu$ is historically called Ioffe time~\cite{Ioffe:1969kf,Braun:1994jq}~\footnote{The term light-cone distance is preferred by some, as it is the light-like distance traveled by the struck quark during the high-energy process.}. In Euclidean space and based on the conventions we use here, $\nu=z P_3$.  $\mathcal{M}(\nu,z^2)$ are called Ioffe-time pseudo-distributions (pseudo-ITDs). One of the main differences between the quasi- and pseudo-distributions formulations is that the former are the Fourier transform of $\mathcal{M}(P_3,z)$ in $z$, while the latter are the Fourier transform of $\mathcal{M}(\nu,z^2)$ in $\nu$. Similarly to quasi-distributions, the pseudo-distributions approach relies on the factorization of pseudo-ITDs obtained on the lattice, to extract the light-cone ITDs, using a matching kernel calculable in perturbative QCD. The necessary matching is performed at the level of ITDs~\cite{Radyushkin:2018cvn,Zhang:2018ggy,Izubuchi:2018srq,Radyushkin:2018nbf} in coordinate space, while in the quasi-PDFs approach is done in the momentum space.

The analysis within the pseudo-ITDs framework is based on a ratio of matrix elements, namely
\begin{equation}
\label{eq:reduced}
\mathfrak{M}(\nu,z^2) = \frac{\mathcal{M}(\nu,z^2)\,/\,\mathcal{M}(\nu,0)}
{\mathcal{M}(0,z^2)\,/\,\mathcal{M}(0,0)},\,
\end{equation}
which are called reduced Ioffe time pseudo-distribution (pseudo-ITD). Such a ratio is a convenient choice and defines a gauge-invariant renormalization scheme. The renormalization scale in pseudo-ITD is set by $1/z$. This consists a major practical difference with the renormalization prescription of quasi-PDFs, which is based on RI$'$-type schemes~\cite{Constantinou:2017sej,Alexandrou:2017huk}. $\mathfrak{M}(\nu,z^2)$ also reduces the higher-twist contamination, as discussed in Ref.~\cite{Orginos:2017kos}. The success of the approach is based on calculating $\mathfrak{M}(\nu,z^2)$ for a wide range of $\nu$, including small-$\nu$ values, by varying either $P_3$, or $z$, or both. Therefore, physical information is included in matrix elements from low values of $P_3$. This is in contrast with the quasi-PDFs, where matrix elements at different $P_3$ are not analyzed together. 

The $\mathfrak{M}(\nu,z^2)$ are matched to light-cone ITDs, $Q(\nu,\mu^2)$ using a matching procedure, and at the same time are converted to the $\overline{\rm MS}$ and evolved to a common scale, $\mu=1/z'$~\cite{Radyushkin:2018cvn,Zhang:2018ggy,Izubuchi:2018srq,Radyushkin:2018nbf}. Finally, the light-cone PDFs are related to $Q(\nu,\mu^2)$ via a Fourier transform in Ioffe time.
\begin{equation}
\label{eq:PDF2ITD}
Q(\nu,\mu^2) =\int_{-1}^1 dx \, e^{i\nu x} q(x,\mu^2),
\end{equation}
Note that this is also subject to the ill-defined inverse problem. One can utilize the crossing symmetries of the distributions, which for the unpolarized case, $q(x)$, reads $\bar{q}(x)=-q(-x)$, where $x>0$. The real part of Eq.~(\ref{eq:PDF2ITD}),
\begin{equation}
\label{eq:ReQ}
{\rm Re}[Q(\nu,\mu^2)] =\int_0^1 dx \cos(\nu x) q_v(x,\mu^2)\,,
\end{equation}
is related to the valence distribution, $q_v=q-\bar{q}$, while the imaginary part,
\begin{equation}
\label{eq:ImQ}
{\rm Im}[Q(\nu,\mu^2)]= \int_0^1 dx \sin(\nu x) q_{v2s}(x,\mu^2)\,,
\end{equation}
to $q_{v2s}\equiv q_v+2\bar{q}=q+\bar{q}$, which is the non-singlet distribution involving two flavors. Therefore, the full and sea-quark distributions can be extracted by combining the two equations. Whether this is possible in practice, depends on the statistical precision of the lattice data. As we will see below, most of the calculations focus on the valence distributions, namely, the real part of $Q(\nu,\mu^2)$.

\begin{figure}[h!]
\begin{center}
\resizebox{0.45\textwidth}{!}{\includegraphics{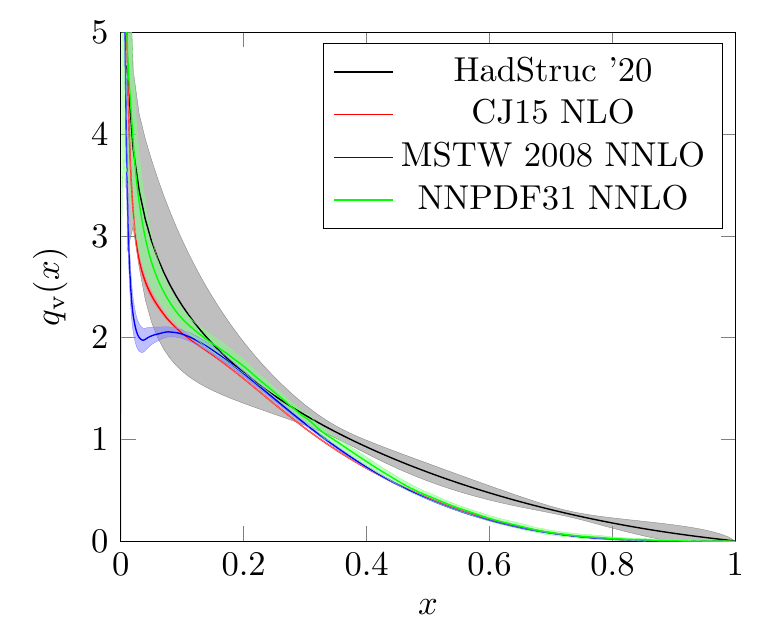}}
\caption{The nucleon valence distribution  compared to phenomenological determinations from the NLO global fit CJ15~\cite{Accardi:2016qay} (green), and the NNLO global fits MSTW2008~\cite{Martin:2009iq} (red) and NNPDF31~\cite{Ball:2017nwa} (blue). Source: Ref.~\cite{Joo:2020spy}. Reprinted based on the arXiv distribution license.}
\label{fig:pdf_extrap}
\end{center}
\end{figure}

The pseudo-ITDs approach has been studied for the unpolarized PDF for the pion and the nucleon in a series of publications, with the more recent ones presented in Refs.~\cite{Joo:2019bzr,Joo:2019jct,Joo:2020spy}. Also, Ref.~\cite{Karpie:2019eiq} discusses the challenges of reconstructing the light-cone distributions from lattice data due to the ill-defined inverse problem, and the need for advanced reconstruction methods. This is relevant for all methods discussed in this article. Here, we focus on the results from Ref.~\cite{Joo:2020spy} on the nucleon valence unpolarized PDF, which uses three ensembles with pion mass 172, 278, and 358 MeV to perform an extrapolation to the physical point. Using the 172 MeV ensemble they calculate the first two moments of PDFs and pseudo-ITDs. The results for $\langle x \rangle$ from pseudo-ITDs and PDFs are in agreement with phenomenological estimates~\cite{Accardi:2016qay,Martin:2009iq,Ball:2017nwa}, while there is some tension for $\langle x^2 \rangle$ using the PDFs lattice estimates for $z^2>4 a^2$. A better agreement is observed for the pseudo-ITDs estimates and NNPDF31~\cite{Ball:2017nwa}. It was checked that higher-twist effects are negligible in accordance with Ref.~\cite{Orginos:2017kos}, as there is no dependence of the moments on $z^2$ within the reported uncertainties. The pion mass dependence between the 172 MeV and 278 MeV is found to be within uncertainties. The same holds for the 358 MeV ensemble, which is, however, much more noisy. Therefore, non-negligible pion mass dependence between data at the physical point and at pion mass above 300-350 MeV is not excluded. Such a dependence was found in Ref.~\cite{Alexandrou:2018pbm}. The final results use a linear extrapolation applied to the lightest two ensembles, and is compared to the phenomenological fits in Fig.~\ref{fig:pdf_extrap}. The lattice data are in agreement with the determinations of Refs.~\cite{Accardi:2016qay,Martin:2009iq,Ball:2017nwa} up to $x \sim 0.25$. Beyond that, the lattice results overestimate the global fits, which is consistent with the tension observed in $\langle x^2 \rangle$. This is because the numerical values of high moments receive important contributions from the large-$x$ region. An in-depth investigation of systematic uncertainties is necessary to address the tension, such as discretization effects and volume effects, which is a multi-year effort and requires multiple ensembles.

\begin{figure}[h!]
\begin{center}
\resizebox{0.48\textwidth}{!}{
\hspace*{-0.25cm}\includegraphics{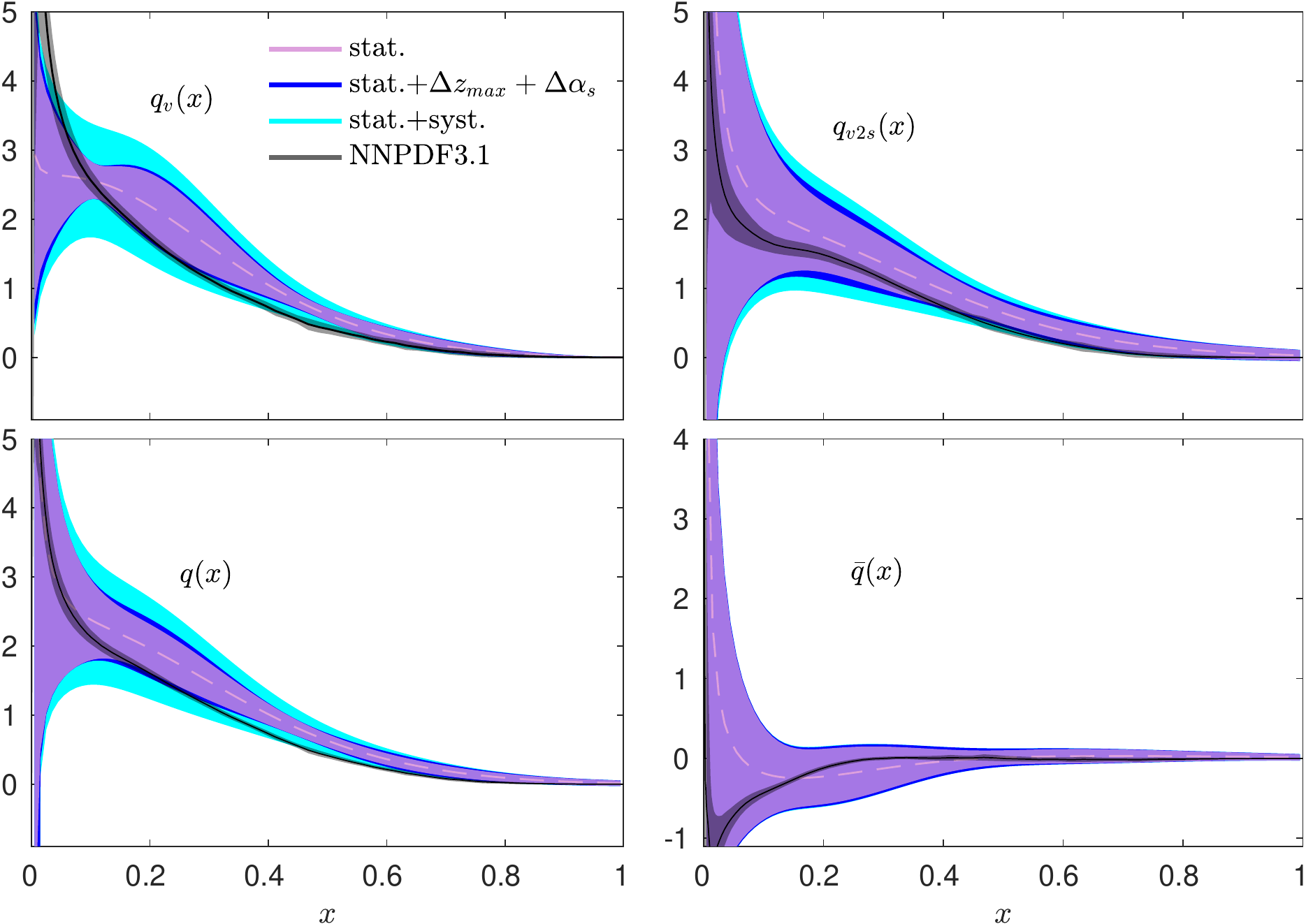}}
\caption{Lattice estimates for the unpolarized PDFs for the valence ($q_v$; upper left), valence + 2 sea ($q_{v2s}=q_v+2\bar{q}$; upper right), full ($q=q_v+\bar{q}$; lower left) and sea ($\bar{q}$; lower right). The global fits of NNPDF \cite{Ball:2017nwa} are shown with a grey band. The bands in the lattice data represent: the statistical uncertainty (purple), the combination of statistical and systematic due to the choice of $\nu_{\rm max}$ and $\alpha_s$ (blue), and the total error including also an estimate for the uncertainties related to cutoff effects, finite-volume effects, excited states contamination, truncation and higher-twist effects (cyan). Source: Ref.~\cite{Bhat:2020ktg}. Reprinted based on the arXiv distribution license.}
\label{fig:final}
\end{center}
\end{figure}

Another recent calculation of the isovector unpolarized PDF using the pseudo-ITDs approach is presented in Ref.~\cite{Bhat:2020ktg}. For the first time, the full PDF, $q(x)=q_v(x)+\bar{q}(x)$, and sea-quark PDF, $q_s(x)=\bar{q}(x)$, are accessed through the calculation of both the valence, $q_v$, and the combination $q_v+2\bar{q}$. The latter is obtained from Eq.~(\ref{eq:ImQ}), which is more difficult to extract, as one requires control over statistical uncertainties. One ensemble at the physical point is used, with $N_f=2$ twisted-mass fermions and lattice extent 4.5 fm ($48^3\times96$). Six values of $P_3$ are used, $P_3=0 - 1.38$ GeV, combined with all possible values of $z$, so that $\nu_{\rm max}\sim 8$. Based on the scale dependence of the reduced pseudo-ITD, it is concluded that $z$ values up to $8 a=0.75$ fm are safe to be used, as the effect is within the reported uncertainties. An extensive investigations is performed on identifying the optimal maximum value of $\nu$ used to extract the final PDFs. It is found that $\nu_{\rm max} = 5.2$ ($z_{\rm max}=0.75$ fm) is optimal. One of the factors to consider is the number of data entering the reconstruction of PDFs given in Eq.~(\ref{eq:PDF2ITD}). One seeks for reliable and statistically meaningful results. Three approaches are explored and compared, namely, a discretized FT, the Backus-Gilbert (BG)~\cite{BackusGilbert}, and the fitting reconstruction (FR) of Ref.~\cite{Joo:2019jct}. The latter is implemented using an ansatz that has the proper small- and large-$x$ behavior, inspired by the global fits of experimental data sets. One of the main differences between the discretized FT and the other methods, is that the former underestimates the uncertainties, as the uncertainties do not reflect the lack of data in certain regions of $\nu$. On the contrary, the BG and FR, have larger uncertainties, which reflects the challenges of the reconstruction, and the amount of information in the lattice data. It is found that the fitting reconstruction performs better than the other approaches, and is, thus, used in the final analysis, as shown in Fig.~\ref{fig:final}.
These data correspond to $\nu_{\rm max}=5.2$ ($z_{\rm max}=0.75$ fm).
The lattice data include both statistical and systematic uncertainties. The various bands shown are due to varying $\nu_{\rm max}$ and the strong coupling. A third band is due to a combined estimate of cutoff effects, finite-volume effects, excited states contamination, truncation and higher-twist effects. The lattice results for $q_v(x)$, $q_{v2s}(x)$, $q(x)$, and $\bar{q}(x)$ are shown and compared to the global fits of NNPDF \cite{Ball:2017nwa}. A remarkable agreement is found for all distributions at all regions of $x$. 

\begin{figure}[h!]
\begin{center}
\resizebox{0.45\textwidth}{!}{\includegraphics{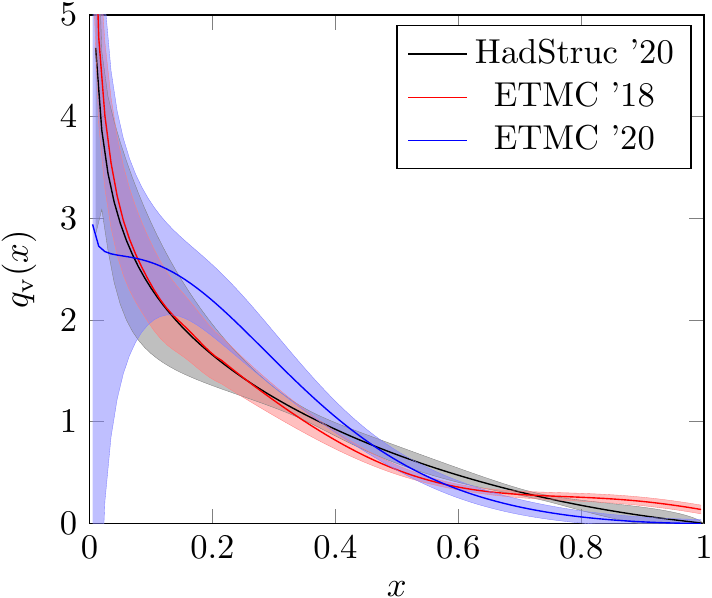}}
\caption{Lattice data for the unpolarized PDF using quasi-PDFs~\cite{Alexandrou:2019lfo} (red band) and pseudo-ITDs from Ref.~\cite{Joo:2020spy} (gray band) and Ref.~\cite{Bhat:2020ktg} (blue band). Source: Ref.~\cite{Joo:2020spy}.  Reprinted with permission of the Authors.}
\label{fig:nucleon_QPFG_pPDF}
\end{center}
\end{figure}

The unpolarized PDF is studied with the same ensemble at the physical point using the quasi-PDFs~\cite{Alexandrou:2019lfo} and the pseudo-ITDs~\cite{Bhat:2020ktg} formulation. A near-physical-point ensemble is also used in Ref.~\cite{Joo:2020spy}, which is used in an extrapolation to the physical point. These three results are compared in Fig.~\ref{fig:nucleon_QPFG_pPDF}. It should be noted that the results of Ref.~\cite{Alexandrou:2019lfo} use the standard discretized FT, while the results of Refs.~\cite{Joo:2020spy,Bhat:2020ktg} are obtained with the fitting reconstruction. Also, the pseudo-ITDs analysis utilizes matrix elements at all values of $P_3$, as they enter the reduced pseudo-ITD and the analysis is in terms of $\nu$. This fact, together with the fitting reconstruction, reduces uncertainties related to the ill-defined inversion problem. Despite the above differences, the results exhibit a very good agreement. The data from quasi-PDFs have a mild oscillation between $x=0.7$ and $x=1$, and are slightly above the results of the other two calculations. This level of agreement suggest that systematic uncertainties related to the inverse problem, to $P_3$ and the maximum value of $z P_3$ are within the uncertainties. One can also address the concern on the renormalization prescription. As mentioned above, the scheme in pseudo-ITDs is gauge invariant and avoids the need of an additional renormalization function in an intermediate RI$'$-type scheme prior the matching to the light-cone PDFs. Furthermore, there are indications that the RI-type scheme is meaningful in the small-$z$ region, $z=3 - 4$ fm~\cite{Ji:2020brr}. Therefore, it is expected that there are systematic uncertainties related to the renormalization of quasi-PDFs. While this is valid, the agreement observed between the results using different lattice formulations and different approaches, suggests that such effects are probably within the reported uncertainties.

\subsubsection{Current-current correlators}
\label{sssec:LCSs}

Besides the matrix elements of bilinear non-local operators used in the quasi-PDFs and pseudo-ITDs methods, there is a more general set of hadronic matrix elements, the good ``lattice cross sections'' (LCSs). The latter can be used to extract PDFs, and eventually analyzed together, in an analogous way as the experimental data~\cite{Ma:2014jla,Ma:2014jga,Ma:2017pxb}. In such global fits of lattice data, systematic effects coming from different results, may have less influence in the combined PDFs. However, this is an ideal scenario and it is not clear how systematic uncertainties from each lattice data will effect the combined fits in practice. Good candidates to construct the set of LCSs share similar features: \textbf{a.} are calculable in lattice QCD; \textbf{b.} are renormalizable; \textbf{c.} have the same and factorizable logarithmic collinear divergences as PDFs. As discussed in Ref.~\cite{Ma:2017pxb}, both the quasi-PDFs, pseudo-ITDs fall under the LCSs category.

A useful choice for the operator is a two-current local operator, $N J_1(z)J_2(0)$, with the currents being space-like separated with distance $z$. Therefore, this operator is gauge invariant, without the need of a Wilson line. $N$ is a dimensionful normalization factor. The matrix elements of such operators (current-current corelators) are renormalized with the multiplication of the two renormalization functions of the local currents, $J_1$ and $J_2$. Therefore, the renormalization procedure is easy and uses well-established techniques. 

A recent work on the pion valence quark distribution using the current-current correlators method is presented in Ref.~\cite{Sufian:2020vzb}. The calculation is performed on $N_f=2+1$ clover fermion ensembles with three values of the pion mass 278, 358, 413 MeV, with two volumes for the heavy pion mass (3 fm and 4 fm). The vector and axial currents are used as $J_1$ and $J_2$, respectively. The momentum boost employed is between 0.41 - 1.65 GeV. Several values of the current separation length are used, up to $z=6a=0.76$ fm. 
As discussed in Ref.~\cite{Sufian:2020vzb}, the signal decays significantly with increasing $P_3$, particularly for $m_\pi=268$ MeV. For example, for current separation of $3a$, the errors are 3 times larger for $P_3=1.65$ GeV as compared to $P_3=0.41$ GeV. Consequently, the data for source-sink separation greater than $12 a=1.5$ fm and $P_3\ge 1.25$ GeV, are not reliable due to the uncontrolled statistical uncertainties. This conclusion is compatible with the findings of other calculations (see, e.g., Sec.~\ref{sec:Discussion}). Only the lattice data with signal-to-noise ratio greater than 1 enter the excited-states analysis. The data from all ensembles are simultaneously analyzed to obtain the functional form of the ITDs. A chiral, continuum, volume, and higher-twist extrapolation is employed using the $z$-expansion, which is frequently used to fit the momentum transferred square, $Q^2$, dependence of form factors. We note that three pion masses, two lattice spacings, and two volumes enter the fit, making the extraction of some of the fit parameters susceptible to systematic uncertainties. Only $z<0.56\,{\rm fm} < 1/\Lambda_{\rm QCD}$ is used in the final analysis, for which the short-distance factorization seems to hold, and the higher-twist contributions are suppressed. However, a negligible dependence on $z$ is found within the reported uncertainties, which is in agreement with the findings of Refs.~\cite{Joo:2020spy,Bhat:2020ktg}. This corroborates the suggestion that the short-distance factorization (SDF) can be applied beyond $0.3 - 0.4$ fm for the current data, as the corresponding systematic effects are within the uncertainties.
Next-to-leading (NLO) order factorization is applied to the extrapolated current-current correlators, and are then parameterized to $q_{\rm v}^\pi(x)$. Two sets of the parameters are identified as good estimates. The analysis shows that larger values of the Ioffe time is needed to better constrain the fit parameters. Given that the values of $z$ must be in a perturbative region, one should increase $P_3$ to achieve higher values of the Ioffe time. As already discussed, increasing $P_3$ while statistical uncertainties are controlled, is a major challenge of the lattice calculations. 
\begin{figure}[h!]
\begin{center}
\resizebox{0.45\textwidth}{!}{\includegraphics{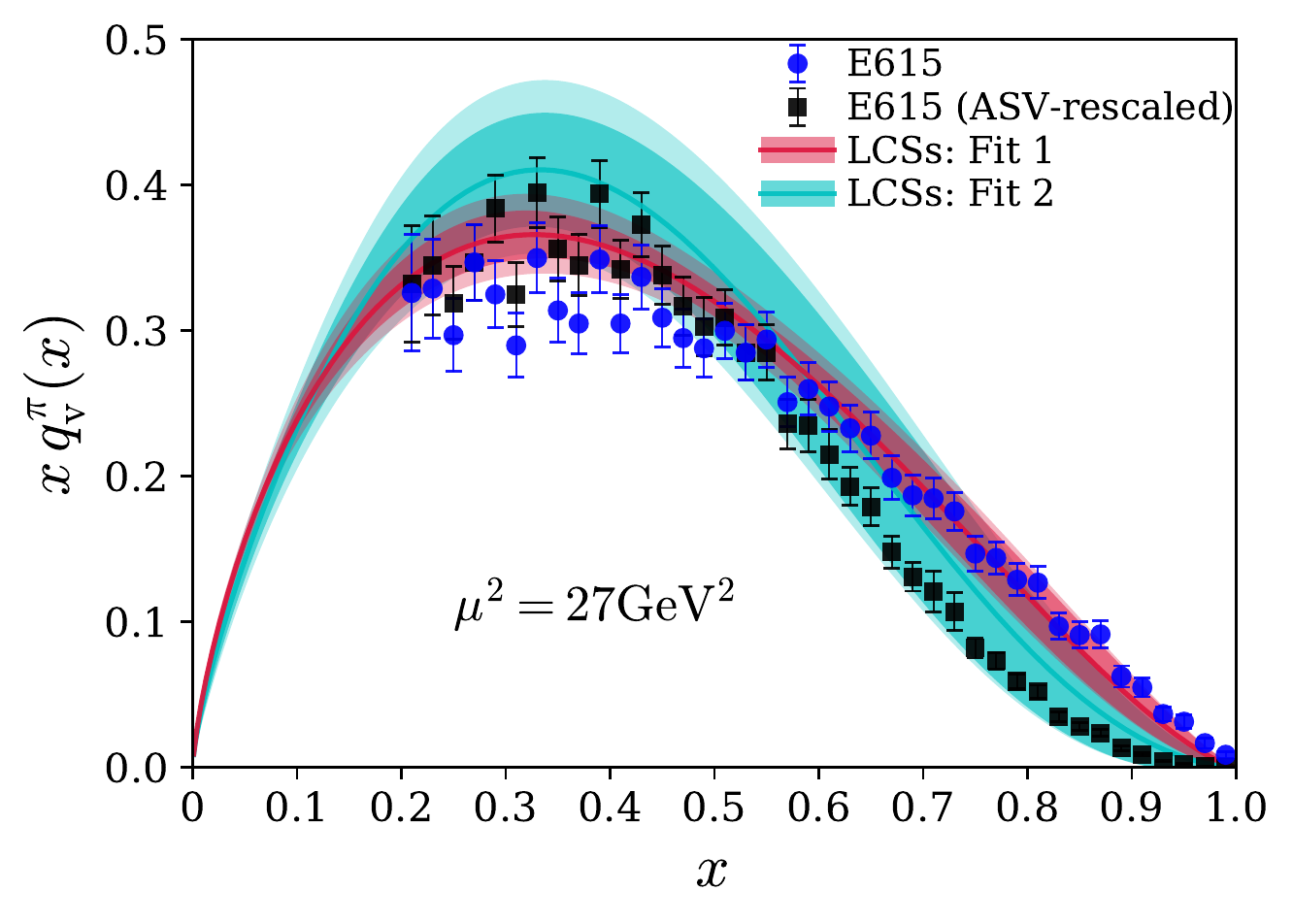}}
\caption{
Lattice data of pion $xq^\pi_{\rm v}(x)$-distribution using two parameterizations (cyan and red bands), shown together with the E615 data from Ref.~\cite{Conway:1989fs} (blue band) and Ref.~\cite{Chang:2014lva} using E615 re-scaled data~\cite{Aicher:2010cb}.  Source: Ref.~\cite{Sufian:2020vzb}. Article published under the terms of the Creative Commons Attribution 4.0 International license.}
\label{fig:evolPDF}
\end{center}
\end{figure}
The lattice-extracted $q^\pi_{\rm v}(x)$ using the two fits are in agreement within errors, as can be seen in Fig.~\ref{fig:evolPDF}. Given the current uncertainties related to the fit, one could consider as a final fit the combination of the two bands. The lattice data are compared with using E156 data~\cite{Conway:1989fs,Aicher:2010cb}. One of the motivations of this work is to identify the large-$x$ behavior of $q_{\rm v}^\pi(x)$, and which functional form it favors. The older analysis of Ref.~\cite{Conway:1989fs} exhibit a $1-x$ fall, while the data of Ref.~\cite{Aicher:2010cb} have a $(1-x)^2$ dependence. Based on the current uncertainties, the lattice data do not favor one over the other. However, the results are very promising, and, in the future, improved data can potentially be used to identify the large-$x$ limit.

\begin{figure}[h!]
\begin{center}
\resizebox{0.45\textwidth}{!}{\includegraphics{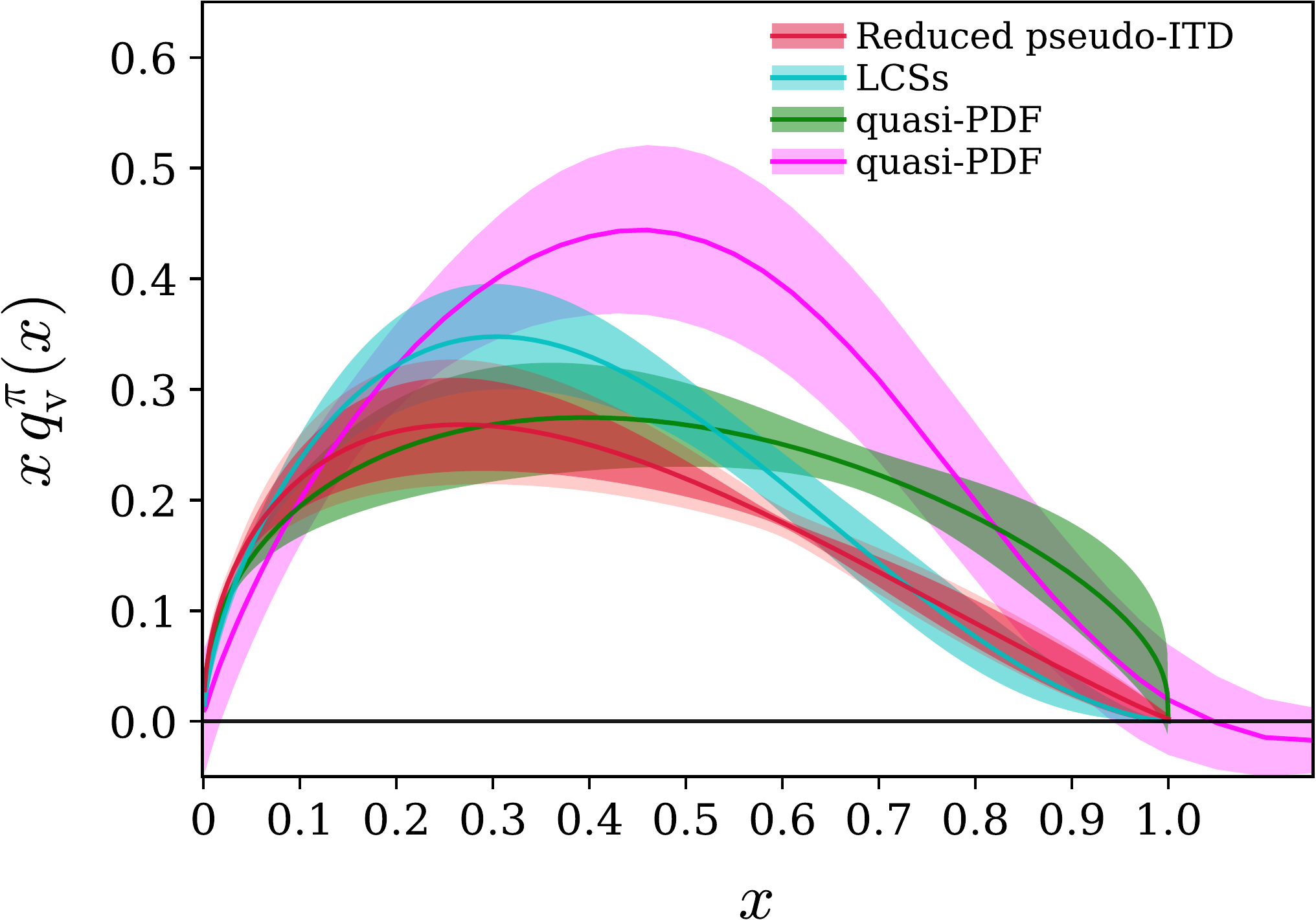}}
\caption{Lattice data for the pion PDF from Ref.~\cite{Chen:2018fwa} on quasi-PDF (pink band), Ref.~\cite{Izubuchi:2019lyk} on quasi-PDF (green band), Ref.~\cite{Joo:2019bzr} on pseudo-ITDs (red band), and Ref.~\cite{Sufian:2019bol} on current-current correlators (cyan band). Source: Ref.~\cite{Joo:2019bzr}. Article published under the terms of the Creative Commons Attribution 4.0 International license.}
\label{fig:pion_QPFG_pPDF} 
\end{center}
\end{figure}

\vspace*{0.25cm}
There is a large number of calculations on the pion distribution using different approaches, lattice formulations, and analysis. It is, therefore, desirable to compare these results. Here we show Fig.~\ref{fig:pion_QPFG_pPDF} taken from Ref.~\cite{Joo:2019bzr}, in which four calculations are compared using quasi-PDFs~\cite{Chen:2018fwa,Izubuchi:2019lyk}, pseudo-ITDs~\cite{Joo:2019bzr} and current-current correlators~\cite{Sufian:2019bol}. 
An $N_f=2+1+1$ mixed-action ensemble of HISQ/clover fermions with pion mass of 310 MeV and volume $24^3\times 64$ is used for the calculation of Ref.~\cite{Chen:2018fwa}. The pion PDF is extracted using the quasi-PDF formulation, and the $x$-dependence reconstruction is performed using the derivative method~\cite{Lin:2017ani}, according to which one neglects the surface term in the integration by parts of the FT. This method does not solve the inverse problem, as discussed in Ref.~\cite{Karpie:2018zaz}. It also introduces further uncertainties (see Sec.~\ref{sec:GPDs}). The results presented in Ref.~\cite{Izubuchi:2019lyk} also employ the quasi-PDFs method and are discussed in Sec.~\ref{sssec:qPDFs}. Instead of a standard FT, two approaches are tested that rely on fits applied to the lattice data in the coordinate space. This is closer to the approach employed in Refs.~\cite{Sufian:2019bol,Joo:2019bzr,Sufian:2020vzb}. The valence pion distribution is shown in Fig.~\ref{fig:pion_QPFG_pPDF}. Note that the various calculations are evolved to different scales: $\mu=2$ GeV~\cite{Joo:2019bzr}, $\mu=3.2$ GeV~\cite{Izubuchi:2019lyk}, and $\mu=4$ GeV~\cite{Chen:2018fwa,Sufian:2019bol}. However, the scale dependence is expected to be small within these values. An agreement is observed between the results of Refs.~\cite{Izubuchi:2019lyk,Sufian:2019bol,Joo:2019bzr} up to $x\sim0.5$, and find the distribution to reach its maximum at around $x=0.3$. Note that Ref.~\cite{Joo:2019bzr} combines two $N_f=2+1$ ensembles with clover fermions reproducing a pion mass of 415 MeV and different volumes, (3 fm and 4 fm). Ref.~\cite{Sufian:2019bol} uses the same ensemble with 4 fm. Even though the two calculations are completely independent and do not share any matrix elements, the distribution extracted is in full agreement. The results of Ref.~\cite{Izubuchi:2019lyk} have a smaller slope after the peak and approaches $x=1$ differently. As mentioned above, a fit on the lattice data is applied to handle the inverse problem, and the $x\to1$ limit is one of the constraints. The results found in Ref.~\cite{Chen:2018fwa} disagree with the other calculations, as they have a maximum around $x=0.5$, and is higher than the other calculations beyond $x=0.5$. It is possible that the use of the derivative method contributes to this disagreement.


\subsubsection{Hadronic tensor}
\label{sec:resultsHT}

The hadronic tensor approach was first proposed in the 90's~\cite{Liu:1993cv,Liu:1998um,Liu:1999ak}, and was recently revived. The methods discussed above rely on factorization (in momentum or coordinate space) and there is a need for large $P_3$ or large $z P_3$ while ensuring that the values used are within the perturbative region. The hadronic tensor method does not require large momentum, as it is frame-independence. It is also free of renormalization. The feasibility of the approach in terms of technical methods can be found in Ref.~\cite{Liang:2019frk}. The method relies on analyzing ratios of suitable four-point and two-point functions, which is computationally very challenging, as one has to suppress the gauge noise and isolate the ground state at the same time. Another difficulty is the reconstructing of the Minkowski hadronic tensor, a process that also suffers from the inverse problem. In-depth investigations are needed to identify which methods are promising.

Ref.~\cite{Liang:2019frk} undertakes such a study using a two ensemble. A mixed action of clover on DWF with $m_\pi=$ 370 MeV is used for the elastic case, $W_{44}$. This serves as the benchmark of the approach, as it should give the vector charge. An anisotropic clover lattice with $m_\pi=$ 380 MeV is used to obtain the Euclidean hadronic tensor with nonzero momentum transfer (3.56 GeV), $W_{11}$. The inverse Laplace transform of the Euclidean hadronic tensor is studied with three methods: the Backus-Gilbert method~\cite{BackusGilbert}, the maximum entropy method~\cite{Asakawa:2000tr}, and the Bayesian Reconstruction method~\cite{Burnier:2013nla}. It is found that the vector charge is reconstructed reliably from $W^E_{44}$, with the maximum entropy method having a better resolution.
\begin{figure}[!h]
\begin{center}
\resizebox{0.5\textwidth}{!}{\hspace*{-0.1cm}
\includegraphics{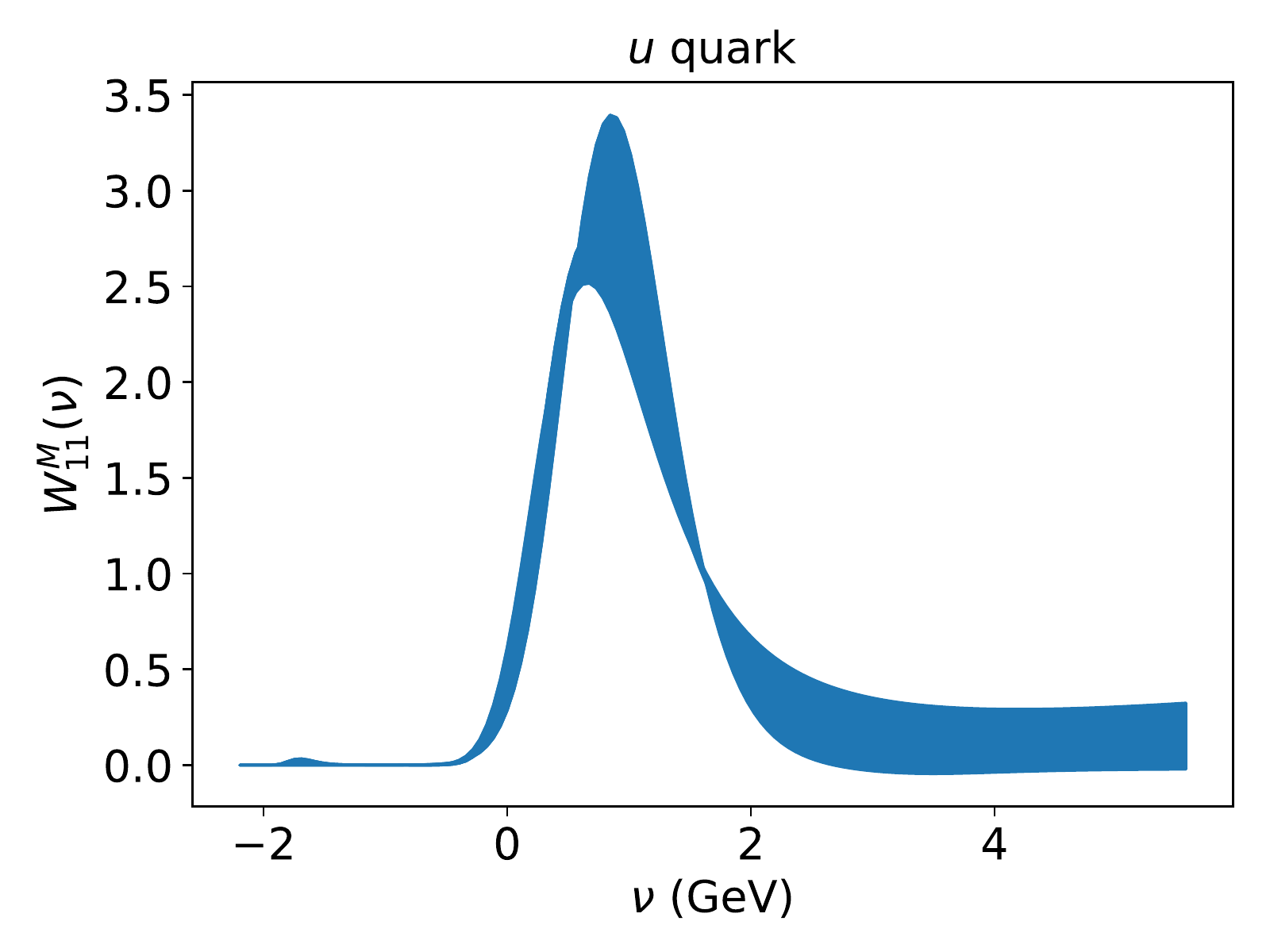}\hspace*{-0.25cm}
\includegraphics{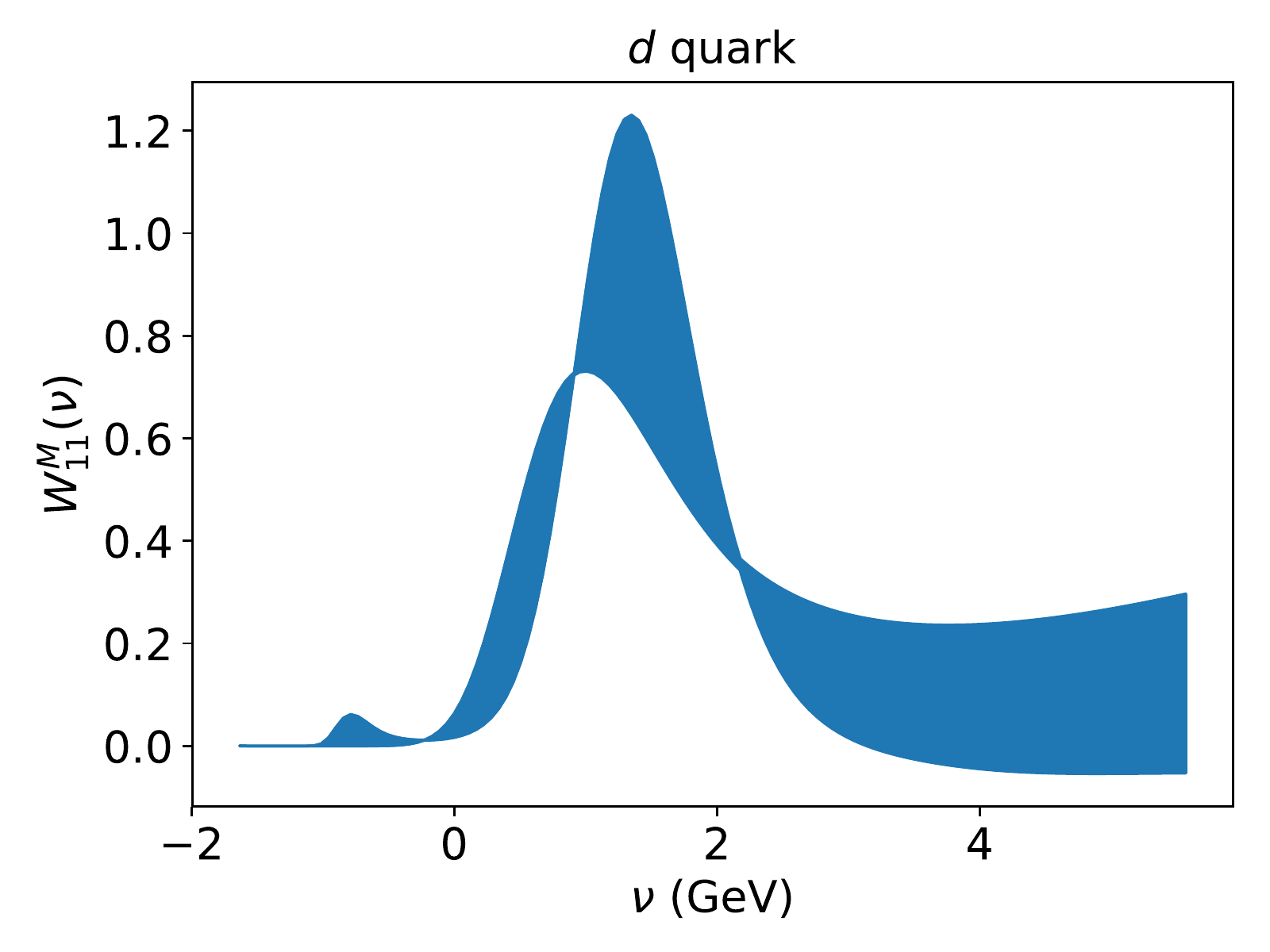}}
\caption{The Minkowski hadronic tensor ${W}^M_{11}$ for $u$-quark (left) and $d$-quark (right) as a function of energy transfer $\nu$. Source: Ref.~\cite{Liang:2019frk}. Article published under the terms of the Creative Commons Attribution 4.0 International license.}
 \label{M2}
  \end{center}
\end{figure}

The reconstructed Minkowski tensor $W_{11}$ in shown in Fig.~\ref{M2} for the up- and down-quark contributions, as a function of the energy transfer. Te reconstruction method used is the maximum entropy method, and the bands represent the uncertainties due to various models. The statistical errors are not considered, possibly because they would dominate the analysis of the the reconstruction methods. Note that the momentum transfer 3.56 GeV is high and the electromagnetic form factors are suppressed. Therefore there is no peak around $\nu=0$, which is the elastic point. There is a peak around $\nu=1$ GeV, which is below the range leading to the parton structure functions ($\nu \in [2.96  - 3.68] \, {\rm GeV}$). The same conclusions are found with the Backus Gilbert method. This is a on-going effort, and improvements are needed in several directions. The ultimate goal is to obtain lattice data for the Minkowski hadronic tensor for a wide set of kinematic setups, which can be used to extract $x$-dependent PDFs via a factorization theorem.

\subsubsection{OPE without OPE}
\label{sssec:OPEwOPE}

Another method to access hadronic distributions was proposed recently in Ref.~\cite{Chambers:2017dov}. It is known as ``OPE without OPE'', as it is related to the work of G. Martinelli~\cite{Martinelli:1998hz}. Based on this approach, one computes matrix elements of the time-ordered product of two local currents,
\begin{equation}
{\cal M}_{\mu\nu}(p,s,s^\prime,z) = \langle p,s^\prime | \,{\cal T} \left\{\mathcal{J}_\mu(z), \mathcal{J}_\nu(0)\right\} \,|p,s \rangle\,.
\end{equation}
The currents are separated by a distance $z$, which has to be small for perturbation theory to be applicable. There is also a lower limit for $z$ for discretization effects to be under control. The Compton tensor 
\begin{equation} \label{eq:compamp}
T_{\mu\nu}(p,q) = \frac{i}{2} \int d^4z\, e^{i q \cdot z} \delta_{s s^\prime} {\cal M}_{\mu\nu} \, ,
\end{equation}
can be decomposed into two scalar functions $\mathcal{F}_{1,2}(\omega,Q^2)$, which are related to the DIS structure functions $F_{1,2}$ from the hadronic tensor decomposition, via the optical theorem:
\begin{equation} 
\label{eq:optical_cuts}
\rm{Im}\mathcal{F}_{1,2}(\omega,Q^2) = 2\pi F_{1,2}(x,Q^2)\,,\quad \omega= \frac{2 p\cdot q}{Q^2}\,.
\end{equation}
The indices $\mu,\,\nu$ control the kinematic factors of the decomposition, with the choice $\mu=\nu=3$ being convenient. Combining Eq.~(\ref{eq:optical_cuts}) with analytical and crossing symmetries, one can write 
\begin{align}
\label{eq:compomega}
\overline{\mathcal{F}}_1(\omega,Q^2)&= 4\omega^2\int_0^1 dx \frac{x\,F_1(x,Q^2)}{1-x^2\omega^2-i\epsilon}\,,\\
\mathcal{F}_2(\omega,Q^2)&= 4\omega \int_{0}^1 dx\,\frac{F_2(x,Q^2)}{1-x^2\omega^2-i\epsilon}\,,
\end{align}
where $\overline{\mathcal{F}}_1(\omega,Q^2) \equiv \mathcal{F}_1(\omega,Q^2) - \mathcal{F}_1(\omega,0)$.

\begin{figure}[!h]
\begin{center}
\resizebox{0.45\textwidth}{!}{\hspace*{-0.1cm}
\includegraphics{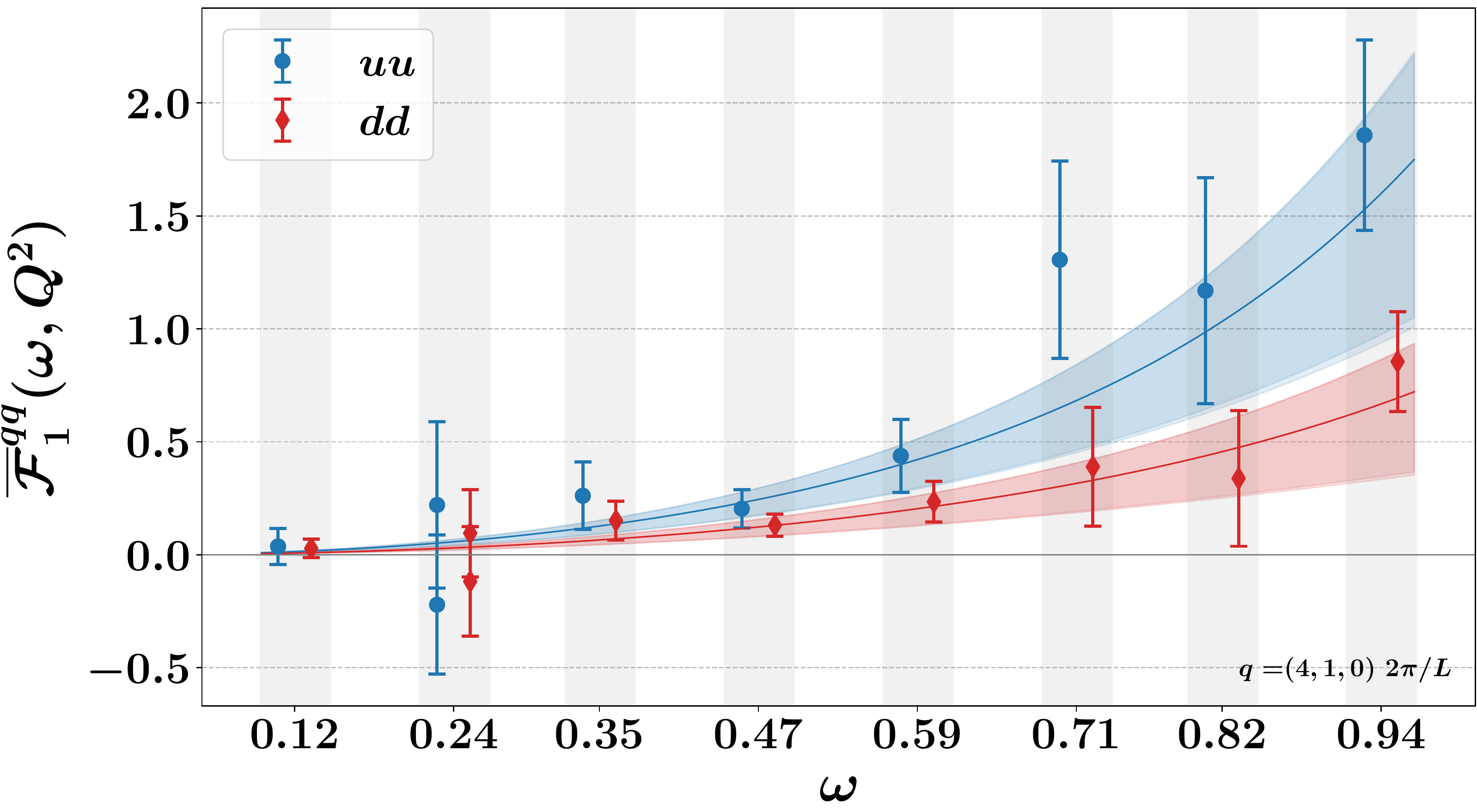}}
\caption{$\omega$ dependence of $\overline{\mathcal{F}}_1^{qq}(\omega, Q^2)$ at $Q^2 = 4.66$ GeV$^2$ for the $uu$ (blue) and $dd$ (red) combinations. Source: Ref.~\cite{Can:2020sxc}. Reprinted based on the arXiv distribution license.}
 \label{fig:q410_moments}
  \end{center}
\end{figure}

The method has been recently studied in Ref.~\cite{Can:2020sxc} using one $N_f=2+1$ ensemble of clover fermions with pion mass 467 MeV and $L=2.4$ fm ($a=0.074$ fm). The matrix element of the current-current operator is computed using the Feynman-Hellmann relation~\cite{Horsley:2012pz}, which avoids the computation of four-point functions. This method requires an additional term in the QCD Lagrangian 
\begin{equation}
\mathcal{L}(x) \rightarrow \mathcal{L}(x) + \lambda \left[ Z_V\cos(\vec{q}\cdot\vec{x})\; e_f \,\bar{\psi}_f(x)\gamma_3 \psi_f(x) \right] \,,
\label{add}
\end{equation}
where $e_f$ is the electric charge of the $f^{\rm th}$ flavor and $\lambda$ is a parameter with dimension of mass. $\vec{q}$ is the external current momentum. To extract $T_{33}$, one needs separate simulations for a few values of $\lambda$, and calculate the derivative of the nucleon energy with respect to $\lambda$, that is
\begin{equation}
\label{eq:FH}
\left. \frac{\partial^2 E_{N_\lambda}({p})}{\partial \lambda^2} \right|_{\lambda=0} = - \frac{T_{33}(p,q) + T_{33}(p,-q)}{2 E_{N}({p})}\,. 
\end{equation}
The flavor combinations considered correspond to the two currents coupled to the same flavor, indicated by $uu$ and $dd$, which are leading-twist. Several values of the current momentum are considered within $3\, {\rm GeV}^2 \lesssim Q^2\lesssim7\, {\rm GeV}^2$. Several values of $p$ and $q$ are used leading to $\omega$ up to 1. The $\omega$-dependence of $\overline{\mathcal{F}}_1(\omega,Q^2)$ for the $uu$ and $dd$ combinations is show in Fig.~\ref{fig:q410_moments}. As can be seen, the quality of the signal is good for both $uu$ and $dd$. Knowledge of $T_{33}(p,q)$ for several values of $\omega$, can lead to the extraction of the $x$ dependence of $F_1$, at fixed values of $q^2$.

\subsection{Flavor decomposition of PDFs}
\label{ssec:singletPDFs}

One of the recent highlights is the calculation of disconnected contributions of matrix elements of non-local operators, which have been analyzed within the quasi-PDFs framework. This is an important direction, as some of the PDFs are not well-constrained from global analyses. For example, most of the high-energy processes are not sensitive to the strangeness. Therefore, it is difficult to disentangle the strange-quark PDF from the down-quark PDF. This results in large ambiguities in its extraction, and, for example, global analysis of DIS + SIDIS data sets gives different sign for $\Delta s$ for different fragmentation functions~\cite{deFlorian:2007aj,Hirai:2007cx}. This poses limitations on the reliable extraction of the $W$-boson mass and the determination of the CKM matrix element $V_{cs}$~\cite{Aaboud:2017svj,Alekhin:2017olj}. Thus, the lattice calculations are quite timely, and while preliminary, demonstrate their potential. To date, there are only 2 calculations at higher-than-the-physical pion mass, which we briefly summarize here~\cite{Zhang:2020dkn,Alexandrou:2020uyt}. Both calculations apply the non-singlet renormalization functions calculated non-perturbatively, and the mixing with the gluon PDFs has been neglected. It should be noted that, this mixing arises at the matching level, and not at renormalization, because there is no additional non-local ultraviolet divergence in the quasi-PDF~\cite{Green:2017xeu,Zhang:2018diq,Wang:2019tgg}. The effects from neglecting both the singlet renormalization and the mixing are expected to be within the uncertainties reported in the calculations. 

Ref.~\cite{Zhang:2020dkn} presents the strange and charm unpolarized PDFs using an $N_f=2+1+1$ ensemble with mixed fermion action (clover on HISQ) with volume $24^3\times64$ ($\sim$3 fm). The nucleon is boosted with momentum up to $P_3=2.18$ GeV. Two values of the valence pion mass are used, that is 310 and 690 MeV, on which a 2-parameter chiral extrapolation is performed, making it less reliable. Also, unlike the 2-parameter fit applied in Ref.~\cite{Joo:2020spy}, here the lightest ensemble is above 300 MeV, possibly not capturing properly the pion mass dependence near the physical point. Furthermore, similar statistical accuracy is required for all data for an unbiased fit. However, the data obtained for the 690 MeV ensemble are a factor of 2-3 more accurate than the 310 MeV data, at $P_3=1.74$ GeV. The lattice data are presented in coordinate space as a function of $z P_3$, and an inverse matching and inverse FT is applied on the global fits from NNPDF31~\cite{Ball:2017nwa} and CT18~\cite{Hou:2019efy}. 
\begin{figure}[h!]
\begin{center}
\resizebox{0.45\textwidth}{!}{\includegraphics{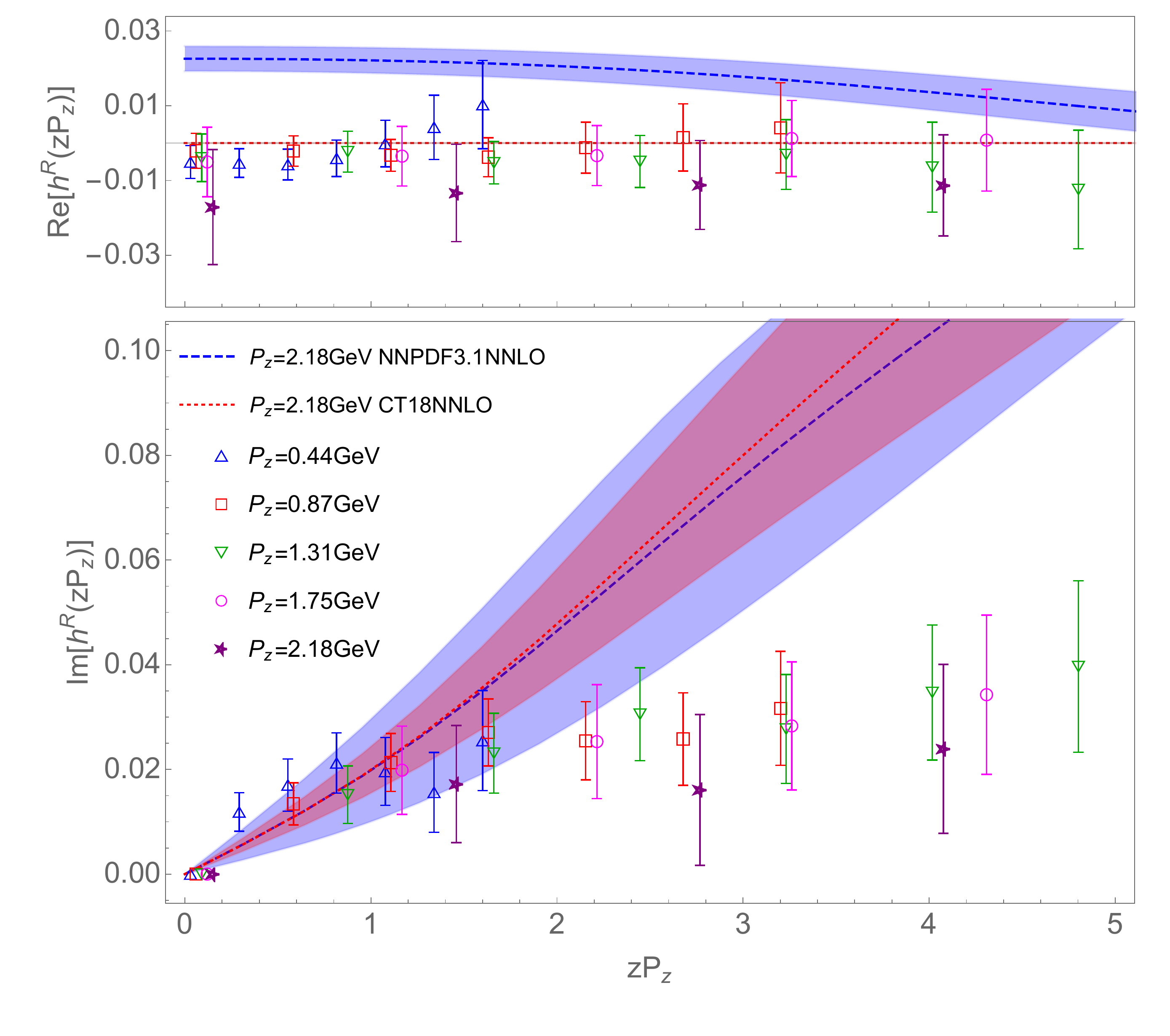}}
\caption{The renormalized unpolarized matrix element (data points) at $P_3$ up to $2.18$ GeV. The global fits of NNDPF3.1~\cite{Ball:2017nwa} and CT18~\cite{Hou:2019efy} are shown upon inverse matching and FT. Source: Ref.~\cite{Zhang:2020dkn}. Reprinted based on the arXiv distribution license.}
\end{center}
\label{fig:lamet_quasi_global_msu}
\end{figure}
The data are shown in Fig.~\ref{fig:lamet_quasi_global_msu} favoring symmetry between $s$ and $\bar{s}$, as the real part of the lattice data is compatible with zero, which is also assumed by CT18. Whether the nonzero signal for NNPDF31 is due to truncation of the matching, is not clear at this stage. The imaginary part is nonzero, and in agreement with CT18 up to $z P_3=12\pi/24 \sim 1.6$, and with the less-accurate NNPDF31 up to $z P_3=18\pi/24 \sim 2.4$. The data shown correspond to $\{n_3,z_{\rm max}\}= \{1, 6a\},\,\{2, 6a\},\,\{3, 6a\},\,\{4, 4a\},\,\{5,3a\}$, where $a P_3=n_3 \times 2\pi/24$, and $z_{\rm max}$ is the maximum value of $z$ used ($a=0.12$ fm). As can be seen, the lattice data for $P_3=1.75$ and 2.18 GeV are very noisy, and thus not very useful. Such rapid decrease of the signal with increase of the momentum boost has been discussed by various groups (see, e.g., Refs.~\cite{Alexandrou:2019lfo,Sufian:2020vzb,Gao:2020ito}). In general, the lattice data have flat behavior, which is not supported by the global fits, demonstrating a tension of a factor of 2 for the large values of $z P_3$. It is assumed that this is due to not taking into account the other flavors in the matching~\cite{Zhang:2020dkn}. Results on the charm unpolarized PDF can be found in Ref.~\cite{Zhang:2020dkn}.

The flavor decomposition of the up- and down-quark PDFs requires the calculation of both the connected and quark disconnected diagram. The disconnected $u+d$ was not calculated until recently, preventing lattice calculations to provide the individual quark PDFs. Such a decomposition is presented for the helicity PDFs, $\Delta u,\,\Delta d$, in Ref.~\cite{Alexandrou:2020uyt}. $\Delta s$ is also presented, which is not combined with the light-quark PDFs, as the mixing with the gluon PDFs is not considered. The calculation is performed using one ensemble of $N_f=2+1+1$ twisted mass fermions, at pion mass 260 MeV, and volume of 3 fm ($32^3\times64$). Three values of $P_3$ are used, with the highest being $1.24$ GeV. Convergence between $P_3=0.83$ GeV and $P_3=1.24$ GeV is observed within the reported uncertainties. Hierarchical probing~\cite{Stathopoulos:2013aci} is used together with momentum smearing, which successfully improves the signal. 
\begin{figure}[h!]
\begin{center}
\resizebox{0.45\textwidth}{!}{\includegraphics{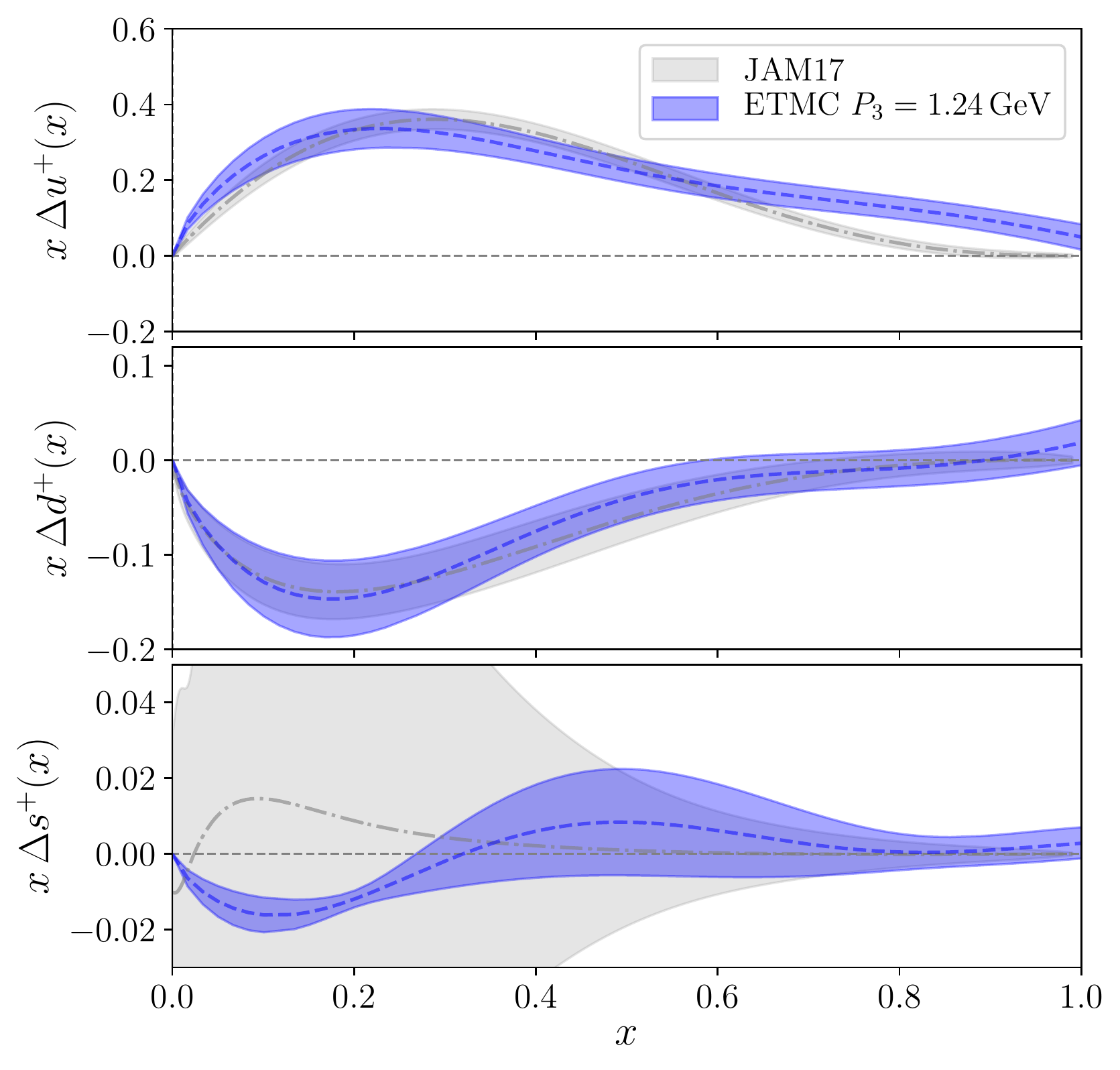}}
    \caption{Comparison of lattice data on the up (upper), down (middle), and strange (bottom) quark helicity PDFs (blue) with the JAM17 phenomenological data~\cite{Ethier:2017zbq} (gray). Source: Ref.~\cite{Alexandrou:2020uyt}. Reprinted based on the arXiv distribution license.}
    \label{fig:flavor_decomposition}
\end{center}
\end{figure}

The final PDFs, $x \Delta^+_q \equiv x\Delta_q + x \Delta \bar{q}$, are shown in Fig.~\ref{fig:flavor_decomposition} for the three flavors, and are compared to the latest global fits of the JAM Collaboration (JAM17)~\cite{Ethier:2017zbq}. Despite neglecting the mixing with the gluon PDFs, the agreement between the lattice results and global fits is very encouraging. $x \Delta^+_u$ exhibits an agreement up to $x\sim 0.65$, while $x \Delta^+_d$ is in full agreement for all $x$ regions. The statistical uncertainties of the lattice data for the light-quark PDFs are comparable to the uncertainties of the global fits. The most striking case, is that of $x \Delta^+_s$, for which the lattice data exhibit smaller statistical uncertainties than JAM17 for $x<0.6$, and both data are compatible within uncertainties for the whole $x$-region. We remind the Reader that the comparison is qualitative, as a thorough investigation of systematic uncertainties is needed, and the elimination of the mixing with the gluon PDFs.

\subsection{Gluon PDFs}

The formalism to extract the $x$-dependence of quark distributions is also applicable for gluons PDFs. This has already been discuss in several papers for the quasi-PDFs~\cite{Wang:2017qyg,Wang:2017eel,Zhang:2018diq,Li:2018tpe,Wang:2019tgg}, pseudo-ITDs~\cite{Balitsky:2019krf}, and current-current correlators~\cite{Ma:2017pxb} methods. \textit{Ab initio} calculations of gluon PDFs is a new direction and largely unexplored. However, it is an essential one because gluonic contributions are sizeable, for example in the proton spin~\cite{Yang:2016plb,Alexandrou:2016ekb,Alexandrou:2017oeh,Yang:2018bft,Alexandrou:2020sml}. Global analysis find that gluon PDFs are dominant in the production of Higgs boson and heavy quarkonium~\cite{Harland-Lang:2014zoa,Butterworth:2015oua,Accardi:2016qay,Hou:2016nqm,Alekhin:2017kpj,Ball:2017nwa,Hou:2019efy,Sato:2019yez}. In fact, the recent analyses constrained better the small-$x$ region, using recent data on the $t\bar{t}$-quark data at the LHC~\cite{Khachatryan:2015uqb,Aaboud:2016pbd,Khachatryan:2016kzg} and Tevatron~\cite{Aaltonen:2015cra}, as well as the HERA run I+II data~\cite{Abramowicz:2015mha}. The domination of gluon PDFs over quark PDFs, especially in the small-$x$ region is a testament of the non-perturbative QCD dynamics responsible for the rich and complex structure of hadrons. Besides their essential role in the hadron structure, calculations of gluon PDFs serve another important purpose: they are necessary to eliminate the mixing with the quark singlet PDFs~\cite{collins_1984}. Disentanglement of the gluon and quark singlet PDFs will eliminate one of the sources of systematic uncertainties in lattice calculations. For the case of Mellin moments of PDFs, the mixing appears in the renormalization. However, as discussed above, the non-local operators are multiplicatively renormalizable~\footnote{This is not generally true for all gluon operators, as some choices mix under renormalization and should be avoided~\cite{Zhang:2018diq,Li:2018tpe}.}. The mixing between the flavor-singlet quark and gluon PDFs should be eliminated at the factorization level.

A preliminary calculation of matrix elements of non-local gluonic operators is presented in Ref.~\cite{Fan:2018dxu}. A mixed action is used with pion mass 330 MeV for the sea sector, and volume 2.7 fm ($24^3\times 64$). The valence sector reproduces a pion mass of 340  and 678 MeV. The nucleon is boosted with $P_3=0.46,\,0.92$ GeV. One of the operators employed exhibits mixing with other gluonic operatros under renormalization, in addition to the mixing with the quark singlet PDFs. Also, the calculation lacks renormalization, and instead, the matrix elements are normalized. Such a normalization changes the quantity under study, and the matching kernel of quasi-PDFs is not applicable. Therefore, only matrix elements are extracted, demonstrating the challenges of the calculation: for $m_\pi=678$ MeV at $P_3=0.92$ GeV the noise exceeds $40\%$ of the signal beyond $z=3a$, while for $m_\pi=340$ MeV the noise is $40\%$ already at $z=0$ and the signal is lost at $z=4a$.

A calculation using the pseudo-ITDs formulation is presented in Ref.~\cite{Fan:2020cpa}, using the same ensemble as Ref.~\cite{Zhang:2020dkn}. The analysis follows Ref.~\cite{Balitsky:2019krf}, for an operator that is multiplicatively renormalizable and a matching kernel known to one-loop accuracy. From the technical standpoint, the advantage of pseudo-ITDs is the cancellation of the renormalization functions in the reduced pseudo-ITD. As discussed in Sec.~\ref{ssec:twist2PDFs}, the RI-type renormalization procedure does not pose a difficulty for quark PDFs, which can be seen by the agreement of unpolarized PDFs obtained from the same ensemble using the quasi- and pseudo-ITDs methods~\cite{Alexandrou:2019lfo,Bhat:2020ktg} (Fig.~\ref{fig:nucleon_QPFG_pPDF}). However, the calculation of the renormalization function for non-local gluon operators cannot be reliably extracted with the methods used for $\langle x \rangle_g$~\cite{Yang:2018bft,Alexandrou:2020sml}. The mixing between the quark-singlet and gluon PDFs appears at the matching procedure for both quasi- and pseudo-ITDs. In Ref.~\cite{Zhang:2020dkn}, the mixing is not considered, as the $u+d$ PDFs for the disconnected diagram have not been calculated. The calculation employs momentum boost from zero up to 2.16 GeV, and consider up to $z=5a = 0.6$ fm, for all momenta. It is found that the matrix elements from different $n_3$ and $z$ at the same $z P_3$ are compatible within the reported uncertainties, even though there are indications that a cut should be applied at around $z=0.33$ fm~\cite{Ji:2020brr}. 
\begin{figure}[h!]
\begin{center}
\resizebox{0.45\textwidth}{!}{\includegraphics{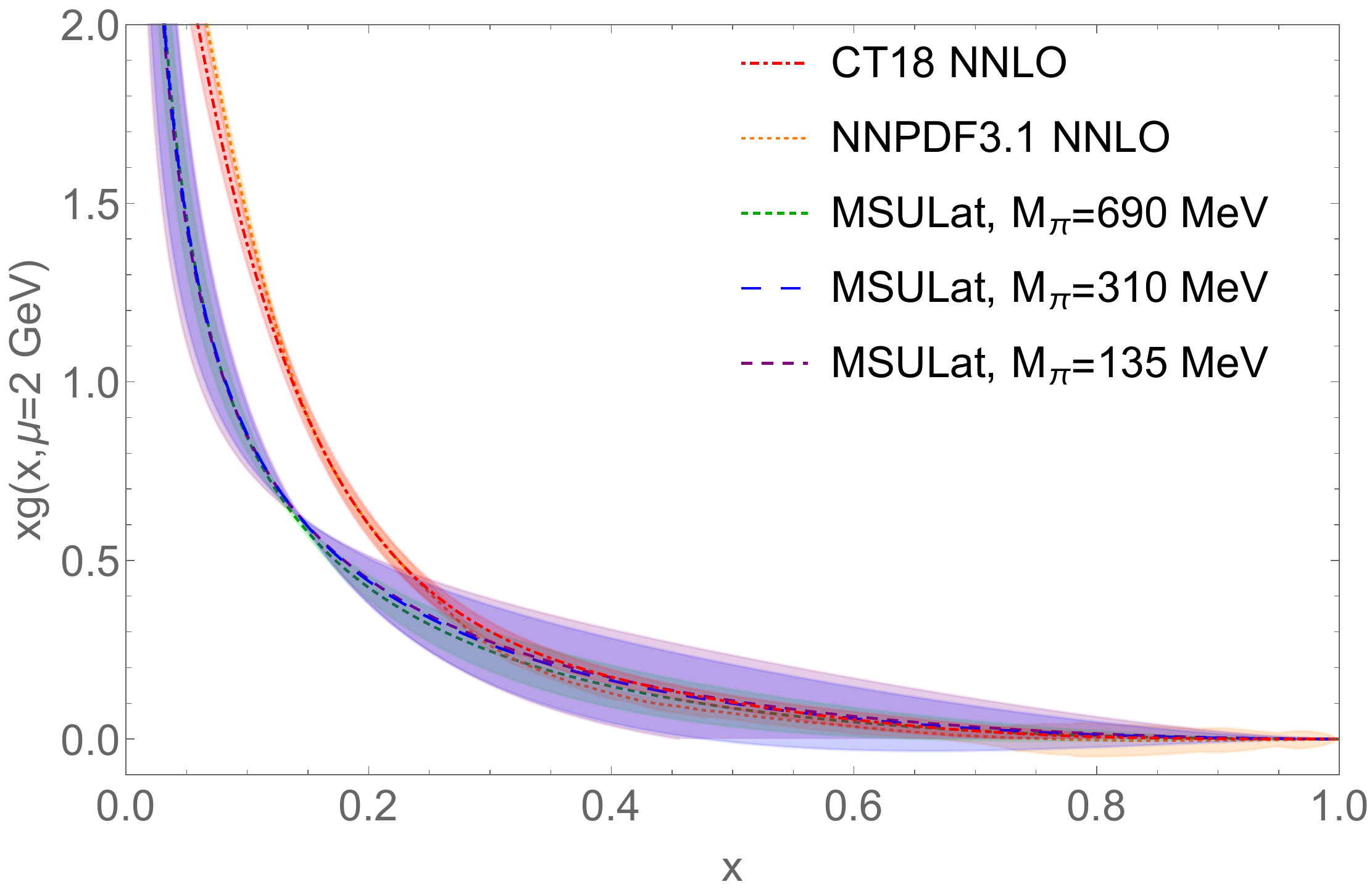}}
    \caption{Lattice data on the unpolarized gluon PDF for $m_\pi=$ 310 and 690~MeV (blue and green bands), and the extrapolated values at $m_\pi=135$ MeV (purple bands). The global fits of CT18 (red band) and NNPDF31 (orange band) are also shown. Reprinted based on the arXiv distribution license.}
\label{fig:xg_PDF}
\end{center}
\end{figure}
It is found that the uncertainties for the 310 MeV ensemble are significantly larger than the corresponding ones at 690 MeV. This happens already at $P_3=1.3$ GeV, where uncertainties of $\sim 30 \%,\, 45 \%,\, 70 \%$ are found for $z=3,\,4,\,5$, respectively, and further increase for the highest momenta. Therefore, only the first few values of $z P_3$ have controlled uncertainties for large $P_3$ in the 310 MeV ensemble, which demonstrates the severe challenges to reliably extract the gluon PDFs.
The final results are shown in Fig.~\ref{fig:xg_PDF}, for the two ensembles. The same 2-parameter extrapolation of Ref.~\cite{Zhang:2020dkn} is also applied here. As discussed in Sec.~\ref{ssec:singletPDFs}, the uncertainties of such procedure are uncontrolled, as only two ensembles are available, with much different statistical uncertainties. The global fits from NNPDF31~\cite{Ball:2017nwa} and CT18~\cite{Hou:2019efy} are also shown for comparison, and an agreement is reported for $x>0.25$. These results are encouraging, but a careful analysis is needed, as the current comparison is inconclusive.

Preliminary results on the gluon pseudo-ITDs using a different lattice formulation have also been presented recently~\cite{Khan}. For results on the gluon helicity $\Delta G$ using LaMET, see Ref.~\cite{Yang:2016plb}.

\subsection{Twist-3 PDFs}
\label{ssec:twist3}

Twist-3 PDFs appear in QCD factorization theorems for a variety of hard scattering processes together with their leading twist (twist-2) counterparts. While twist-3 PDFs lack a probabilistic interpretation, they are interesting at their own right, as they contain information on the soft dynamics, such as the transverse polarization. Also, they are not smaller in magnitude than the twist-2 PDFs, and thus, are important to be considered. Most of the work on twist-3 PDFs is theoretical (see e.g., Refs.~\cite{Balitsky:1987bk,Burkardt:2001iy,Kanazawa:2015ajw,Pitonyak:2016hqh,Aslan:2018zzk}), while the experimental studies are limited. It is very challenging to probe them experimentally and isolate them from the leading-twist contributions. In addition, their theoretical analysis is more evolved, and goes beyond the simple partonic model. A recent re-analysis of the JLab Hall A data on the absolute DVCS beam helicity cross section differences~\cite{Defurne:2016eiy,Defurne:2017paw} addressed twist-3, but it is inconclusive on their contribution. 

Twist-3 PDFs that enter the description of transverse-spin observables in inclusive high-energy processes, include contributions from a number of matrix elements. The latter are classified as intrinsic, kinematical, and dynamical (see, e.g., Ref.~\cite{Kanazawa:2015ajw}). Intrinsic twist-3 observables are extracted from matrix elements with two-parton non-local operators, the kinematical ones are first moments of transverse momentum dependent (TMD) two-parton PDFs, and the dynamical ones from matrix elements with quark-gluon-quark non-local operators.

The most-well studied structure function is the transverse spin dependent $g_2^{\rm s.f.}$~\cite{Jaffe:1996zw}, from which one can separate the non-perturbative component, the intrinsic $g_T$ PDF. The properties of $g_2^{\rm s.f.}$ in DIS were first studied by Wandzura and Wilczek~\cite{Wandzura:1977qf}, proposing the Wandzura-Wilczek (WW) approximation. In their model calculation, it is suggested that the Mellin moments of $g_T(x)$ receive contributions from twist-2 operators and twist-3 operators, allowing one to write
\begin{equation}
g_T(x) = g_T^{\rm WW}(x) + g^{\rm twist-3}_T(x)\, ,
\end{equation}
where $g^{\rm twist-3}_T$ the contribution from twist-3 operators~\cite{Accardi:2009au}. In the WW approximation~\footnote{The WW-approximation has been discussed mostly for the structure function $g_2^{\rm s.f.}$, but the same arguments hold for the corresponding parton distributions.}, the moments of twist-3 operators vanish, and thus, $g_T(x)$ is determined only by the twist-2 operator of the helicity PDF $g_1(x)$. The validity of the WW-approximation has been under investigation since its proposal. In Ref.~\cite{Accardi:2009au} the WW approximation was tested using experimental data, leading to possible violation at the level of $15-40\%$.

The field of extracting the $x$-dependence of PDFs using the methodologies outlined in Sec.~\ref{sec:PDFs} has reached sufficient maturity that enables the study of twist-3 PDFs and their properties. To date, there is only one lattice calculation
\footnote{Several years ago, a calculation on the second moment of $g_T(x)$, $d_2$, was studied using several ensembles with pion mass greater than 550 MeV~\cite{Gockeler:2005vw}. The results were compared to experimental data from the Jefferson Lab Hall A Collaboration~\cite{Flay:2016wie}, finding encouraging agreement.}  
on twist-3 PDFs, in particular, on $g_T(x)$ presented in Ref.~\cite{Bhattacharya:2020cen}. The calculation is performed using $N_f=2+1+1$ twisted mass fermions at a pion mass of 260 MeV, and employing the quasi-PDFs method. The relevant matrix element is $\langle P\vert \, \overline{\psi}(0,z)\,\gamma^j\,\gamma^5\, W(z)\,\psi(0,0)\,\vert P\rangle$, where $j$ is in one of the transverse directions.
The maximum proton momentum boost used is $P_3=1.67$ GeV. The necessary matching was calculated in~\cite{Bhattacharya:2020xlt}, demonstrating that the twist-3 light-cone PDFs and their quasi-PDF counterparts have the same infrared physics, as in the case of all twist-2 PDFs. It was also shown that there are no zero-mode contributions. An extension of this work to the twist-3 $e(x)$ and $h_L(x)$ PDFs, has shown that the singular zero-mode contributions cannot be avoided, making the extraction of their matching more complicated~\cite{Bhattacharya:2020jfj}. 
\begin{figure}[h!]
\begin{center}
\resizebox{0.45\textwidth}{!}{\includegraphics{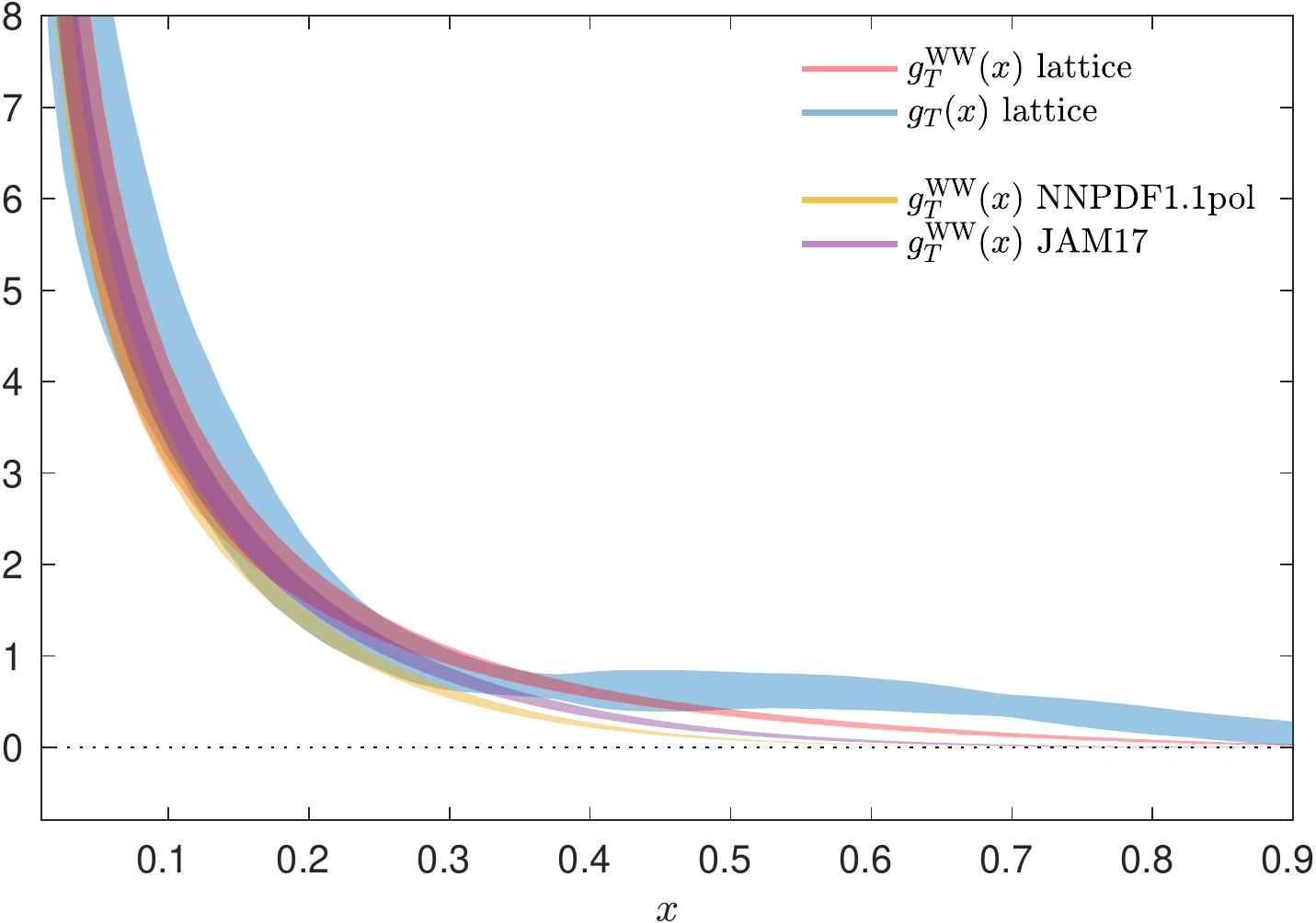}}
	\caption{Comparison of lattice estimate of $g_T(x)$ (blue band) with its WW approximation (red band). The WW approximation using global fits also shown (NNPDF1.1pol~\cite{Nocera:2014gqa} orange band, JAM17~\cite{Ethier:2017zbq} purple band). Source: Ref.~\cite{Bhattacharya:2020cen}. Reprinted based on the arXiv distribution license.}
 	\label{fig:gT_WW}
\end{center} 	
\end{figure}
The lattice data on $g_1(x)$ have been used to test the WW approximation, which suggests that it fully defines $g_T$(x), via
%
\begin{equation}
\label{eq:gT_WW}
g_T^{\rm WW}(x)=\int_x^1 \frac{dy}{y} g_1(y)\,.
\end{equation}
%
The lattice results on $g_T^{\rm WW}(x)$, as well as the actual lattice results on $g_T(x)$ are compared in Fig.~\ref{fig:gT_WW} for $P_3=1.67$ GeV. The estimate from global fits by the NNPDF \cite{Nocera:2014gqa} and JAM \cite{Ethier:2017zbq} collaborations are also shown. As can be seen, the lattice $g_T(x)$ is consistent with the lattice $g_T^{\rm WW}(x)$ for $x \le 0.5$. However, not all sources of systematic uncertainties have been accounted for, and thus, a violation of the WW approximation is still possible at the level of up to 40\% for $x\lesssim0.4$, which is similar to the one found using experimental data~\cite{Accardi:2009au}. The comparison is, of course, qualitative, as simulations at the physical pion mass and investigation of systematic uncertainties is essential. It should also be noted that there is mixing between the intrinsic and dynamical twist-3 operators, which should also be dealt at the matching level, as discussed above for the gluon and singlet quark PDFs. However, twist-3 is a completely unexplored area for the quasi-PDFs approach. While a lot of work is needed in the direction of twist-3 PDFs, the work of Refs.~\cite{Bhattacharya:2020cen,Bhattacharya:2020xlt,Bhattacharya:2020jfj} demonstrate the feasibility of lattice calculations and the matching formulation.

\subsection{Other developments}
\label{ssec:devel}

In parallel to improving the numerical calculations, there have been interesting developments in a more theoretical direction, and here we discuss a few of them. 

Some of the highlights regarding the matching formalism are the calculations presented in Refs.~\cite{Chen:2020arf,Chen:2020iqi,Li:2020xml,Chen:2020ody}, which extend the matching kernels to next-to-next-to-leading order (NNLO). In practice, this means two-loop calculations with respect to the strong coupling constant. It has been discussed extensively that, the truncation of the matching formula to NLO consists one of the major sources of systematic uncertainties. Therefore, this is an important development. In particular, the flavor non-diagonal quark-quark channel in the modified minimal subtraction scheme for the quasi-PDFs is discussed in Refs.~\cite{Chen:2020arf,Chen:2020iqi}. The NNLO matching for the valence PDFs in coordinate space is presented in Ref.~\cite{Li:2020xml}, and the NNLO matching for quasi-PDFs in Ref.~\cite{Chen:2020ody}. The calculations reach a consensus on the importance of the two-loop, emphasizing that this is the natural extension for lattice calculations.

As discussed in previous subsections, there are restrictions in the $z$-range so that factorization is applicable, and this also concerns the renormalization. In an RI-type scheme, for example, there could be non-perturbative effects in the large-$z$ region through logarithms, such as $\log(z^2 \mu^2)$. The pseudo-ITDs ratio which is a renormalization scheme, could also hinder higher-twist effects in large $z$. However, both the RI-type and ratio schemes are suitable to remove all divergence for small and intermediate values of $z$. This consists the main motivation of Ref.~\cite{Ji:2020brr}, proposing a hybrid renormalization, which treats the short-distance and long-distance renormalization separately. This consists of using an RI-type, or ratio scheme up to some $z_{\rm max}$, and a mass-subtraction scheme for the $z$ beyond $z_{\rm max}$. At $z_{\rm max}$, the renormalization functions from the two schemes should be the same based on continuity arguments. It is estimated that an upper limit for $z_{\rm max}$ is around 0.4 fm. It is possible that this constrain for $z_{\rm max}$ is beyond the current precision level of state-of-the-art calculations. For example, the agreement observed in Fig.~\ref{fig:nucleon_QPFG_pPDF} and Fig.~\ref{fig:pion_QPFG_pPDF} between lattice data from different lattice actions and renormalization schemes, does not indicate large effect from treating renormalization differently. The validity of extrapolating the lattice data based on the asymptotic long-range Regge-type behavior~\cite{Regge:1959mz} is also presented. This type of fits are also used in the pseudo-ITDs and current-current correlators method. 

Ref.~\cite{DelDebbio:2020cbz} addresses the renormalization and the validity of the methods factorizing equal-time correlators into light-cone PDFs. The work is based in one-loop calculations of quasi-PDFs and pseudo-PDFs in a renormalizable scalar theory. The calculation addresses the erroneous claims~\cite{Rossi:2017muf,Rossi:2018zkn} that quasi-PDFs cannot be related to light-cone PDFs because the moments of the former exhibit power-divergent mixing. It should be noted that the claims have been refuted in more than one publications~\cite{Ji:2017rah,Radyushkin:2018nbf,Karpie:2018zaz}. For instance, one of the counter-arguments is that, while all moments except for the zeroth do not converge, the light-cone PDFs are extracted from the non-local quasi-distributions that avoid the power divergence problem~\cite{Ji:2017rah}. The non-local operators under study contain logarithmic divergences, as well as power-law divergence due to the Wilson line. Both are renormalizable by adopting a non-perturbative RI-type prescription. In addition, the required matching kernel is responsible for extracting light-cone PDFs with finite moments. A numerical example was given in Ref.~\cite{Karpie:2018zaz} calculating the two lowest moments of pseudo-PDFs, and found agreement with the results of a direct calculation of moments using one- and two-covariant derivative local operators. Finally, the use of a Taylor expansion in $z$ in Refs.~\cite{Rossi:2017muf,Rossi:2018zkn}, is only justifiable when all derivatives with respect to $z^2$ exist at $z{=}0$~\cite{Radyushkin:2018nbf}, which is not the case for the non-local operators discussed here. 
The work of Ref.~\cite{DelDebbio:2020cbz} generalizes Collins' work~\cite{Collins:1980ui} to off-light-cone matrix elements for leading twist and found that quasi-PDFs and pseudo-PDFs are equivalent. These, as well as their light-cone counterparts are renormalizable. It is also demonstrated that factorization exists at small $z$, large $P_3$ and small flow times. Yet again, the claims of Refs.~\cite{Rossi:2017muf,Rossi:2018zkn} have been proven invalid. Light-cone PDFs can indeed be accessed properly  via the methods discussed in this review, as well as other.

\subsection{Synergy of lattice QCD and global fits}
\label{ssec:pheno_lat}

Recently, there has been an interest from the phenomenological community to incorporate lattice data in the global analyses, in the same fashion as the experimental data sets. The main motivation is to explore whether the PDFs can be constrained more successfully, in regions where the experimental data are either sparse, imprecise, or non-existing.

\begin{figure}[h!]
\begin{center}
\resizebox{0.44\textwidth}{!}{\includegraphics{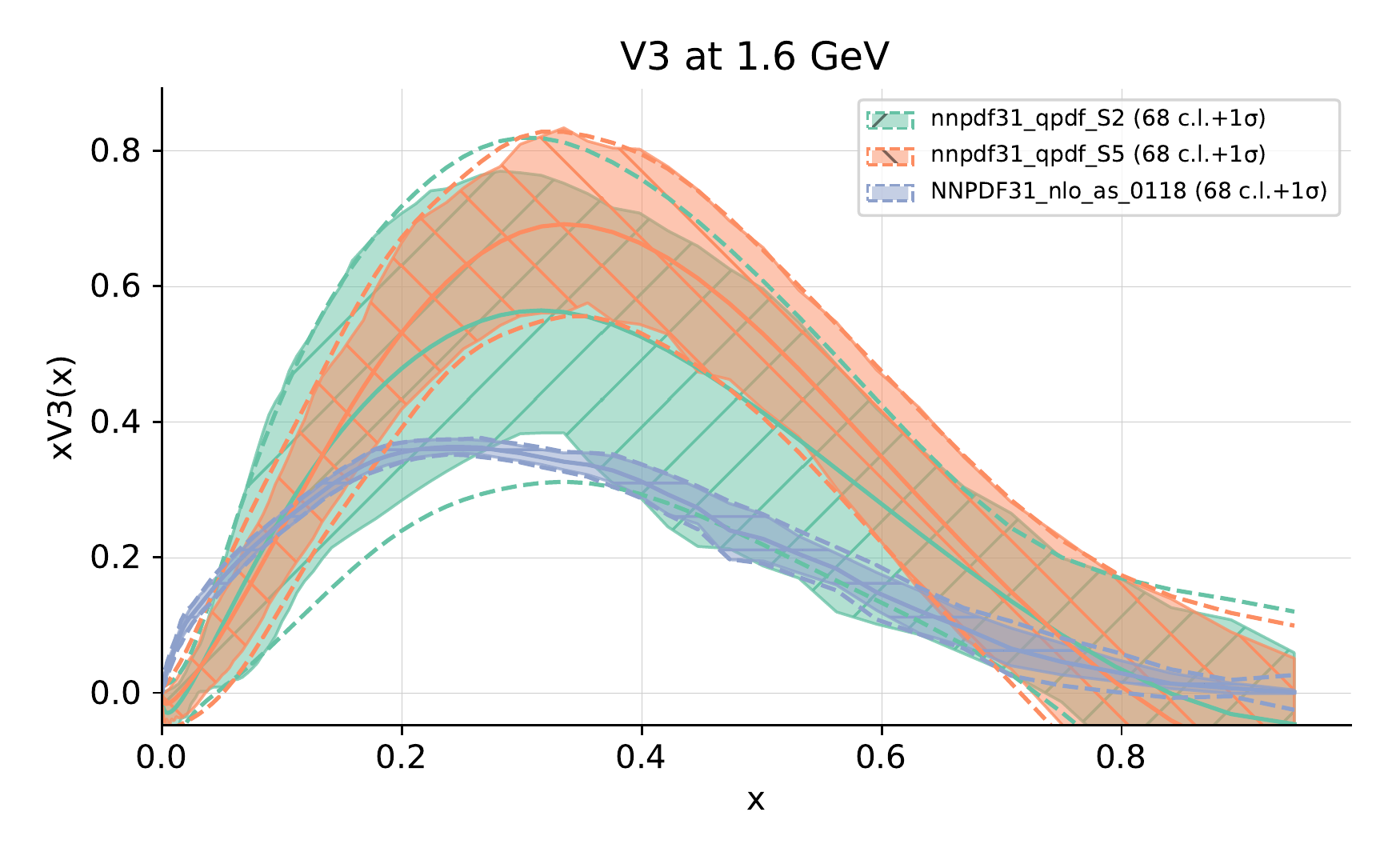}}
\resizebox{0.44\textwidth}{!}{\includegraphics{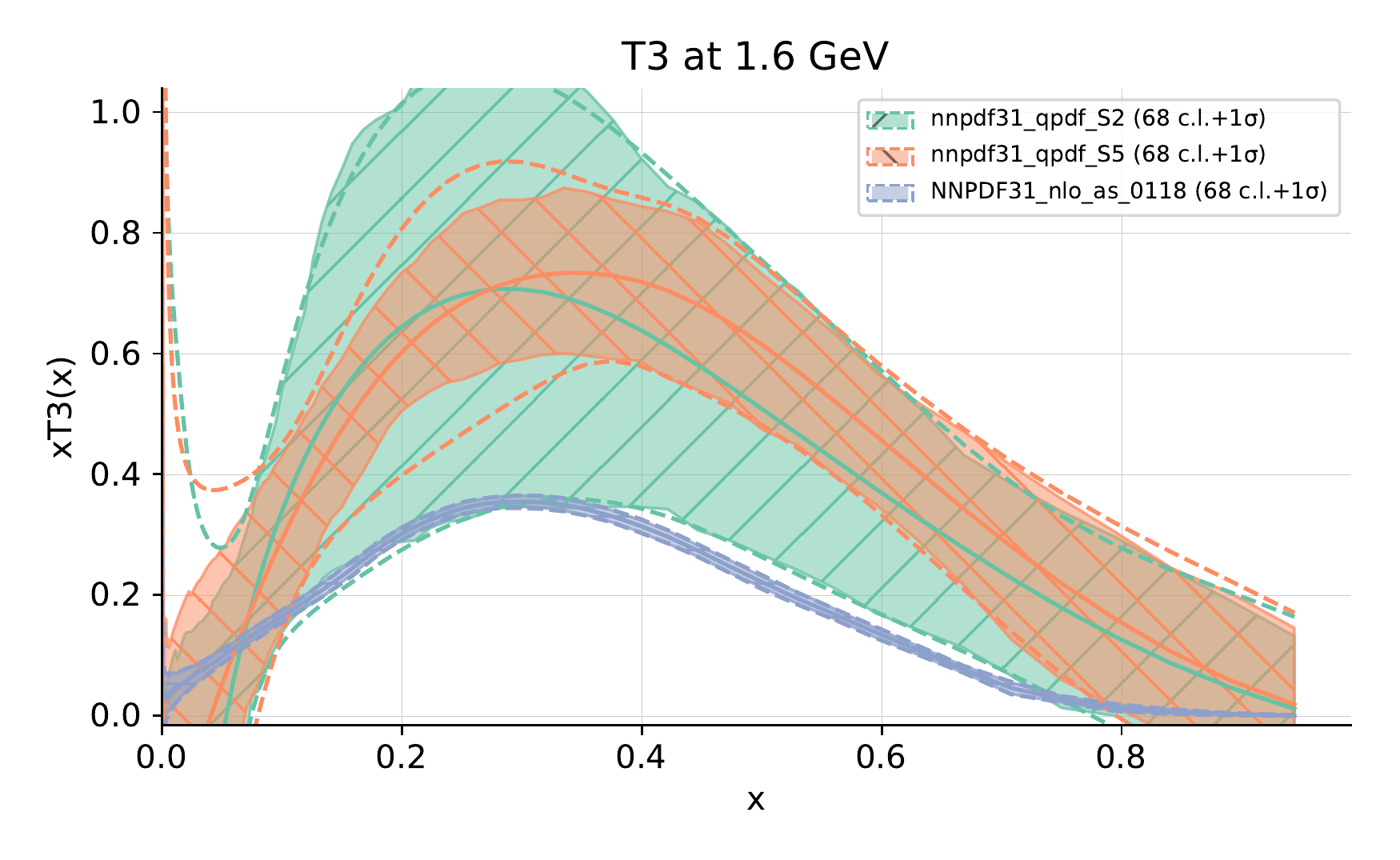}}
\caption{$V_3$ (top) and $T_3$ (bottom) non-singlet PDF using lattice data with systematic uncertainties based on scenario S2 (green band) and S5 (salmon band). The NNPDF31 estimates are shown with a blue band. Source: Ref.~\cite{Cichy:2019ebf}. Published under open access license.}
\label{fig:lattice_data}
\end{center}
\end{figure}
A general framework to extract PDFs from the lattice data treating them on the same footing as experimental data is presented in Ref.~\cite{Cichy:2019ebf}. The methodology is implemented for the unpolarized PDFs within the NNPDF fitting framework using the results of Refs.~\cite{Alexandrou:2018pbm,Alexandrou:2019lfo}. Two non-singlet flavor combinations are considered, that is
\begin{align}
& V_3 ( x ) = u\left(x\right) - \bar{u}\left(x\right) -\left[d\left(x\right)-\bar{d}\left(x\right)\right]\, , \\
& T_3 ( x ) = u\left(x\right) + \bar{u}\left(x\right) -\left[d\left(x\right)+\bar{d}\left(x\right)\right]\, ,
\end{align}
which are related to the real and imaginary parts of the vector matrix element, via
\begin{align}
\label{eq::V3factorization}
\hspace*{-0.1cm}
\text{Re}\left[M(\nu,z^2,\mu)\right] 
&{=} \int_{0}^{1} dx \,\mathcal{C}_3^{\text{Re}}\left( \nu x, z, \frac{\mu}{P_z}  \right) V_3\left(x,\mu\right) \, , \\
\label{eq::T3factorization}
\hspace*{-0.1cm}
\text{Im}\left[M(\nu,z^2,\mu)\right] 
&{=} \int_{0}^{1} dx \,\mathcal{C}_3^{\text{Im}}\left( \nu x, z, \frac{\mu}{P_z}  \right) T_3\left(x,\mu\right)\,.
\end{align}
Therefore, the lattice matrix elements can be analyzed within the framework applied on experimental data sets. The procedure is based on finding a $\chi^2$-function, which is minimized in order to extract $V_3$ and $T_3$. Quantifying statistical and systematic uncertainties is important in this process. Six plausible scenarios are considered to assess systematic uncertainties that are not known, focus on the two realistic ones, labeled S2 and S5. Scenario S2 (S5) attributes 20$\%$, $5\%$, $10\%$, $20\%$ (0.2, 0.05, 0.1, 0.2) for cut-off effects, finite-volume effects, excited-states contamination, and truncation of perturbative expansion, respectively. Note that the systematic uncertainties in S2 change with $z$, while they are fixed in S5. The analysis of lattice data including these systematic effects is shown in Fig.~\ref{fig:lattice_data}, and are compared to the phenomenological fits NNPDF31. It is observed that the more conservative scenario S2 leads to larger uncertainties, and is marginally compatible with the global analysis estimate at most regions of $x$. This analysis emphasizes the importance of properly quantifying systematic uncertainties in lattice estimates.

\begin{figure}[h!]
\begin{center}
\resizebox{0.4\textwidth}{!}{\includegraphics{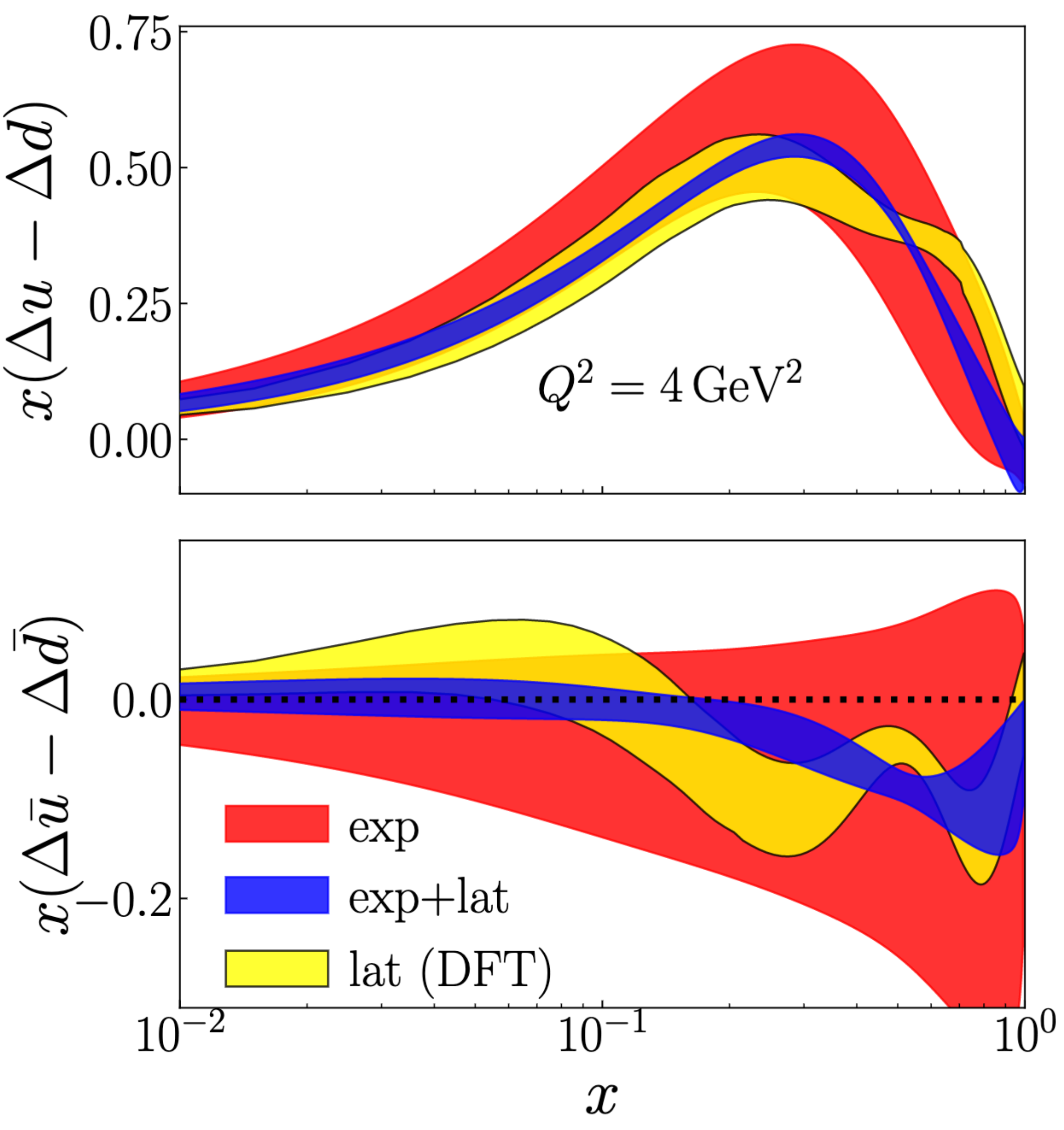}}
\caption{The isovector helicity PDF. The JAM17~\cite{Ethier:2017zbq} results are shown with red band, while the blue band shows the combined fits using both lattice and experimental data. The yellow band shows the data from Refs.~\cite{Alexandrou:2018pbm,Alexandrou:2019lfo} using the standard discretized FT. Source:~\cite{Bringewatt:2020ixn}. Reprinted based on the arXiv distribution license.}
\label{fig:JAMpol}
\end{center}
\end{figure}

Ref.~\cite{Bringewatt:2020ixn} presents a combined Monte Carlo based QCD global analysis within the JAM framework for both the unpolarized and helicity PDFs. The lattice data of Refs.~\cite{Alexandrou:2018pbm,Alexandrou:2019lfo} are treated under the same conditions as the experimental data sets from unpolarized and polarized DIS and Drell-Yan (note this is not the same as the JAM17 results~\cite{Ethier:2017zbq}). The analysis uses the Monte Carlo Bayesian inference sampling methodology, parametrizing the PDFs at the input scale $\mu$, and solving the evolution equations in Mellin space to bring all the observables at the same scale. In a nutshell, the analysis shows a rather weak constraint from the lattice data on the phenomenological unpolarized PDFs. This behavior is in agreement with the findings of Ref.~\cite{Cichy:2019ebf}. In contrast, the lattice data on the helicity PDFs provide significant constraints to the combined estimates, as can be seen in Fig.~\ref{fig:JAMpol}. It is found that the combined estimates of the quark PDFs have reduced uncertainties by a factor of 3 in the small- and large-$x$ regions, and a factor of 5 in the intermediate-$x$ regions. The anti-quark isovector helicity PDF exhibits an equally large improvement. It is also interesting that the combined fit (``exp+lat'') is compatible with the discretized FT (``DFT'') of Refs.~\cite{Alexandrou:2018pbm,Alexandrou:2019lfo}. It should be noted that the implementation of the lattice data is done at the level of the renormalized matrix elements in coordinate space. Therefore, the challenges of the inverse problem do not appear directly in this analysis, as the discretized lattice data are fitted within the JAM framework. However, the effectiveness of such fits are driven by the number and quality of the lattice data. As mentioned previously, properly quantifying both statistical and systematic uncertainties is crucial.

Based on the analyses of Refs.~\cite{Cichy:2019ebf,Bringewatt:2020ixn}, one can conclude that the current precision of the lattice data on the unpolarized case must improve significantly to reach the quality of the experimental data. The case of the helicity PDFs is more promising, as the independent analysis of Refs.~\cite{Alexandrou:2018pbm,Alexandrou:2019lfo} shows a very good agreement with the global fits, for a wide range of $x$. In fact, the lattice data compare favorably with the experimental data sets, and demonstrate a great constraining power, as shown in Fig.~\ref{fig:JAMpol}. The work of Refs.~\cite{Cichy:2019ebf,Bringewatt:2020ixn} opens up new possibilities for synergy between lattice QCD and global QCD fits. It is expected that, in the near future, combined analysis of multiple factorizable lattice matrix elements from various approaches will be available, an idea already discussed in Refs.~\cite{Ma:2014jla,Ma:2017pxb}. Ref.~\cite{DelDebbio:2020cbz} proposes a framework of a global analysis of lattice data.

\vspace*{1.5cm}
\section{GPDs}
\label{sec:GPDs}

GPDs were first introduced in the 1990's, and it was demonstrated that they can be accessed in Deeply-Virtual Compton Scattering (DVCS) and Deeply-Virtual Meson Production (DVMC)~\cite{Mueller:1998fv,Radyushkin:1996nd,Ji:1996ek}. Of particular interest for spin physics, it was shown by Ji~\cite{Ji:1996ek}, that the proton angular momentum can be decomposed in a gauge invariant way into quark and gluon parts. GPDs are relevant to non-forward scattering processes that capture the momentum transfer between the initial and final hadron states. They depend on two additional kinematic variables in addition to $x$, that is, the light-cone component of the longitudinal momentum transfer (skewness), $\xi$, and the momentum transfer squared $t=\Delta^2$. Note that the transverse component of the longitudinal momentum transfer enters through $t$. Consequently, the GPDs are multi-dimensional quantities and mapping their additiona two-dimensional functional surface $(t,\xi)$ allows one to get information on the two-dimensional structure of hadrons, in terms of these kinematic variables.

There are a number of challenges in extracting GPDs from experimental data, as the latter are limited, and cover a small kinematic region. Not only these data sets are indirectly related to GPDs through the Compton form factors, but it is very difficult to disentangle the GPDs entering the same high-energy process~\cite{Diehl:2003ny,Ji:2004gf,Belitsky:2005qn,Kumericki:2016ehc}.

The $x$-dependence of GPDs for fixed values of $t$ and $\xi$, can be accessed using the methods developed for PDFs. This new direction has recently been explored using the quasi-PDFs method in Refs.~\cite{Chen:2019lcm,Alexandrou:2020zbe}~\footnote{The work presented in Ref.~\cite{Lin:2020rxa} does not calculate the matrix element in the symmetric (Breit) frame, which is necessary to obtain the standard, frame-dependent GPDs. See discussion below.} The possibility to access GPDs through the pseudo-GPDs method has also been proposed~\cite{Radyushkin:2019owq}, but yet to be implemented in lattice calculations.

The connected contribution for the pion GPD, $H_v^\pi$, is presented in Ref.~\cite{Chen:2019lcm}. The motivation is to study the large-$x$ behaviour and test which parameterization of kinematic dependence of the GPD is preferable~\cite{Diehl:2004cx,Burkardt:2002hr}. The calculation is performed using an $N_f=2+1+1$ mixed-action ensemble of HISQ/clover fermions with pion mass of 310 MeV and volume $24^3\times 64$. Only zero skewness is considered in the calculation, and two values of the momentum transfer, that is $-t\sim 0.40,\,0.92$ GeV$^2$. Since there is only one GPD for the pion, one avoids the complications of disentangling GPDs, and the need for more than one independent matrix elements to achieve the isolation. As a consequence, the calculation of the pion GPDs is computationally less expensive than the nucleon ones. The derivative method is employed, according to which one neglects the surface term in the integration by parts of the FT. This process is controlled only if the renormalized matrix elements decay to zero. Another concern about this method is that the surface term contains a factor of $1/x$, which makes the extraction of the PDF for small values of $x$ less reliable. In Ref.~\cite{Alexandrou:2019lfo} it was demonstrated numerically that the difference between the discretized FT and the derivative method differ up to $x=0.2$ for the case of the nucleon unpolarized PDFs.
\begin{figure}[h!]
\begin{center}
\resizebox{0.45\textwidth}{!}{\includegraphics{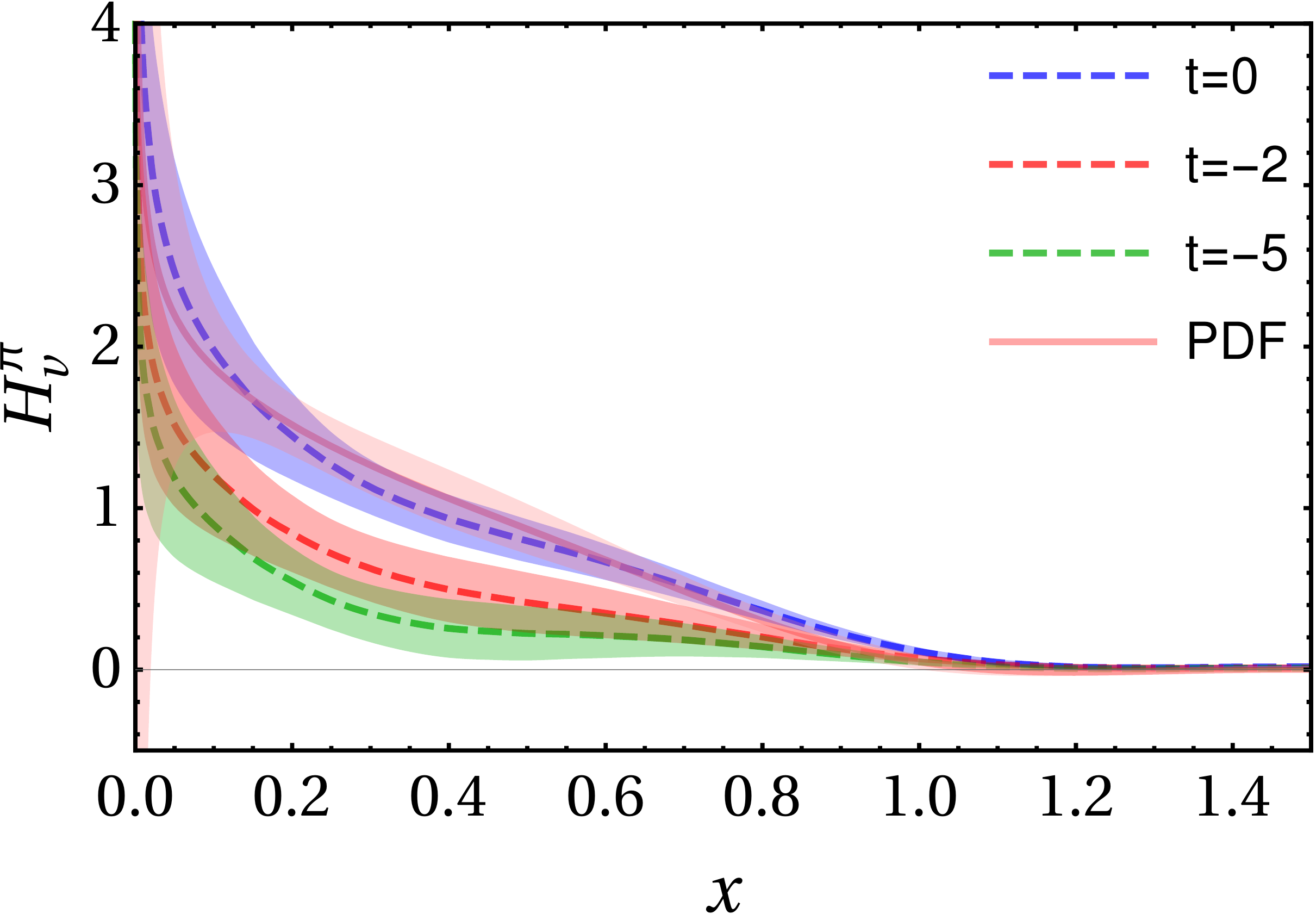}}
\caption{The $x$-dependence of the pion GPD $H^{\pi^{+}}_{v}$ for zero skewness and $t=0.40,\,0.92$ GeV$^2$ (red and green bands), as well as the corresponding PDF (blue and salmon bands). Source: Ref.~\cite{Chen:2019lcm}. Article published under the terms of the Creative Commons CC-BY license.}
\label{fig:xqvpi}
\end{center}
\end{figure}
The main result of the work is shown in Fig.~\ref{fig:xqvpi} for the pion PDF, and GPD for $-t\sim 0.40,\,0.92$ GeV$^2$ at $P_3=1.74$ GeV. Meson-mass corrections have also been applied. One observes that as $t$ increases, the GPD decreases in magnitude. It is concluded that the accuracy of these results is not sufficient to favor one of the two parameterizations, and further studies are necessary.

The isovector unpolarized and helicity GPDs for the proton, $H(x,\xi,t),\,E(x,\xi,t)$ and $\widetilde{H}(x,\xi,t),\,\widetilde{E}(x,\xi,t)$, have been studied using an $N_f=2+1+1$ ensemble of the twisted mass formulation~\cite{Alexandrou:2020zbe}. The ensemble has volume $32^3\times64$ and pion mass 260 MeV. Both zero and nonzero skewness ($\xi=\pm 1/3$) have been studied, for $P_3$ up to 1.67 GeV, and $-t=0.69,\,1.39$ GeV$^2$. As these are discretized on the lattice, specific combinations of the above values of $P_3,\,\xi,\,t$ have been obtained. Note that the skewness defined in the lattice calculation (quasi-skewness) is defined as $\xi=-\frac{Q_3}{2P_3}$ and defers to the light-cone $\xi$ in power corrections. As mentioned above, extracting nucleon GPDs is computationally very costly. In addition, one requires a sophisticated analysis to properly disentangle the GPDs, as each matrix elements contributes, in general, to more than one GPDs.~\footnote{An exception is the helicity case for $\xi=0$, which contributes only to $\widetilde{H}(x,\xi,t)$, as the kinematic factor of $\widetilde{E}(x,\xi,t)$ vanishes.} For example, the off-forward proton matrix element of the vector current decomposes into
\begin{eqnarray}
\label{eq:unpol_decomp}
 \langle N(P_f)|{\cal O}_{\gamma_\mu}(z)|N(P_i)\rangle = \hspace*{4.5cm} \nonumber \\ \langle\langle\gamma_\mu\rangle\rangle F_H(z,P_3,t,\xi) -i\frac{\langle\langle\sigma_{\rho\,\mu}\rangle\rangle\, Q_\rho}{2m} F_E(z,P_3,t,\xi)\,,\hspace*{0.4cm}
\end{eqnarray}
where $Q\equiv P_f-P_i$, and $\langle\langle \Gamma \rangle\rangle \equiv \bar{u}_N(P_f,s')\, \Gamma \,u_N(P_i,s)$ with $u_N$ the proton spinors. Therefore, a minimum of two parity projectors are needed, leading to two independent matrix elements, in order to disentangle $F_H$ and $F_E$. Another critical aspect of the calculation of GPDs is the need for implementation of the Breit (symmetric) frame. In this setup, both the momentum of the initial and final states, $P_i, \, P_f$ carry half of the momentum transfer, that is $\vec{P}_i=P_3 \hat{z} -\vec{Q}/2$ and $\vec{P}_f = P_3 \hat{z} +\vec{Q}/2$. This has major implications in the computational cost, as separate calculations are needed for every value of $t$, because both the source and sink momenta change. We emphasize that there is a fundamental difference between the calculation of Mellin moments of GPDs (form factors and generalized form factors) and the light-cone GPDs. The former are frame independent, and one can extract them in a more convenient frame, that is, fix the sink momentum and attribute $t$ to the source ($\vec{P}_f=P_3\hat{z}$, $\vec{P}_i=P_3 \hat{z} -\vec{Q}$). However, the light-cone GPDs are frame dependent, and their standard definition is in the symmetric frame. The connection between different frames is not a simple kinematical one, and therefore, the lattice calculations must be performed in the frame where GPDs are defined. Introducing momentum transfer in the matrix element, coupled to the momentum boost of the proton states, increases significantly the statistical noise. Another complication of $t\ne0$ is related to the momentum smearing method, which is necessary to suppress the statistical uncertainties. As pointed out in Ref.~\cite{Bali:2016lva}, the exponent of the momentum smearing phase must be parallel to the proton momentum to ensure overlap with the ground state. Therefore, all ingredients of the optimized ratio between 3pt- and 2pt-correlation functions (see, e.g., Eq. (S4) of the supplementary document of Ref.~\cite{Alexandrou:2020zbe}) requires a separate momentum smearing. 

\begin{figure}[h!]
\begin{center}
\resizebox{0.45\textwidth}{!}{\includegraphics{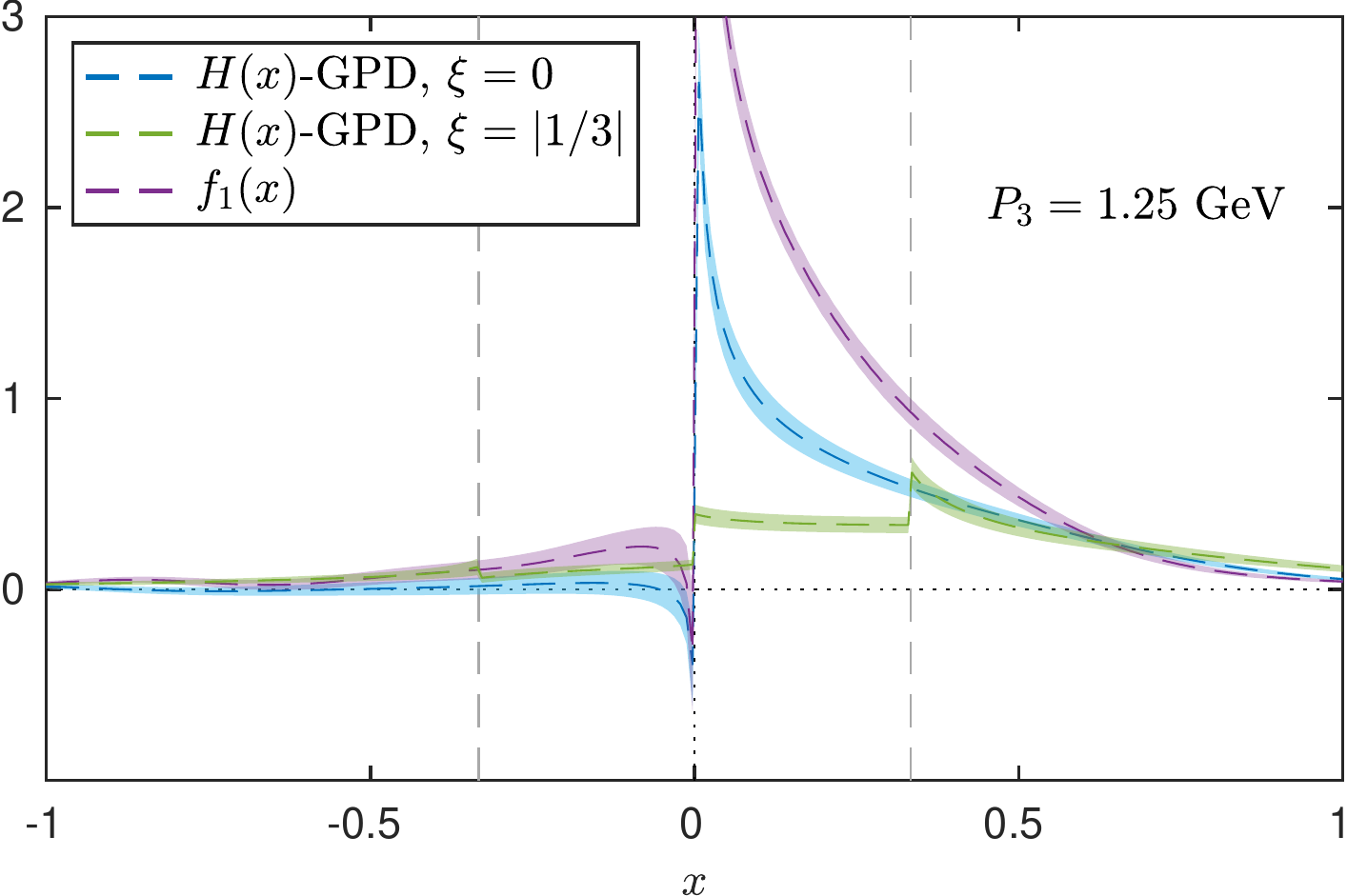}}
\caption{$H(x,\xi,t)$ for $\xi=0$ (blue band) and $\xi=|1/3|$ (green band), and $f_1(x)$ (violet band) for $P_3=1.25$ GeV. The area between the vertical dashed lines is the ERBL region. Source: Ref.~\cite{Alexandrou:2020zbe}. Reprinted based on the arXiv distribution license.} 
\label{fig:H_vs_PDF_p2}
\end{center}
\end{figure}
\begin{figure}[h!]
\begin{center}
\resizebox{0.45\textwidth}{!}{\includegraphics{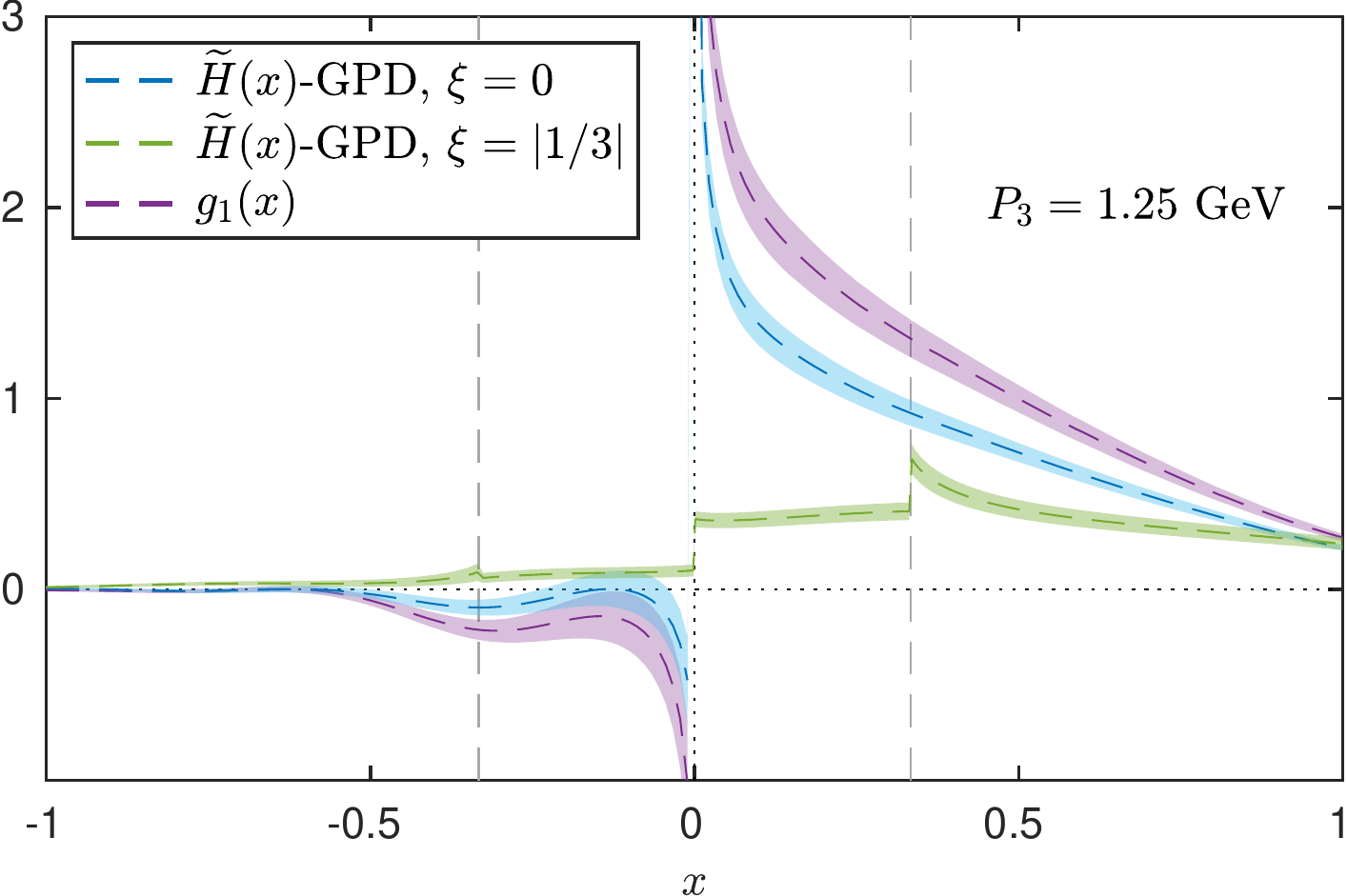}}
\caption{Comparison of $\widetilde{H}(x,\xi,t)$ and $g_1(x)$. Notation as in Fig.~\ref{fig:H_vs_PDF_p2}. Source: Ref.~\cite{Alexandrou:2020zbe}. Reprinted based on the arXiv distribution license.} 
\label{fig:Htilde_vs_PDF_p2}
\end{center}
\end{figure}
Figs.~\ref{fig:H_vs_PDF_p2} - \ref{fig:Htilde_vs_PDF_p2} show $H(x,\xi,t)$ and $\widetilde{H}(x,\xi,t)$ at $P_3=1.25$ GeV, for which both zero and nonzero skewness are available. Each is compared to the corresponding PDF, $f_1(x)$ and $g_1(x)$, respectively. As can be seen, there is a qualitative difference between the GPDs at zero and nonzero skewness. Let us remind the Reader that this mean zero and nonzero momentum transfer in the direction of the boost. In the latter case, the $x$-range separates into two kinematic regions, the ERBL ($|x|< \xi$), and the DGLAP region ($|x|> \xi$), which have different physical interpretation~\cite{Ji:1998pc}. From the practical standpoint, the matching for $\xi \ne 0$ becomes more complicated than the one for $\xi=0$, with the latter being the same as for the PDFs~\cite{Liu:2019urm}. The PDFs are more dominant, as the probability of finding quarks with $t\ne 0$ decreases with increase of $t$. For the unpolarized case, we can compare the lattice data in the DGLAP region with the power-counting analysis of Ref.~\cite{Yuan:2003fs} at the $x\to 1$ limit. One of the predictions of this analysis is that $H(x,\xi=0,t)$ approaches $f_1(x)$, which is confirmed by the lattice data. It also predicts that at $x\to 1$ the $t$-dependence vanishes, which is also observed by the lattice data. For results on the $E$-GPD and $\widetilde{E}$-GPD, as well as results on the highest momentum, see Ref.~\cite{Alexandrou:2020zbe}.

\vspace*{1.5cm}
\section{TMD PDFs}
\label{sec:TMDs}

To map the three-dimensional structure of hadrons, knowledge of TMD PDFs is necessary, as it provides information not only on the longitudinal momentum, but also on the transverse momentum of partons. TMD PDFs arise in processes that involve multiple kinematic scales, such as Drell-Yan, $e^{+} e^{-}$ annihilation, and semi-inclusive deep inelastic scattering. Consequently, a more complicated factorization formalism is required. For instance, the differential cross section is sensitive to the small transverse momenta, for which the non-perturbative dynamics cannot be neglected. Thus, the TMD PDFs contain the small transverse momenta information, making them more detailed than collinear PDFs. 

The TMD factorization of the singular cross-section is not unique, and the various schemes are related to each other. The original proposal for TMD PDFs by Collins, Soper and Sterman~\cite{Collins:1981uk,Collins:1981va,Collins:1984kg} is such that the hard factor is absorbed in the definition of TMDs. In the revised definition for the TMD factorization~\cite{Collins:1350496}, the TMD PDFs have two components: the beam function (one for each incoming hadron) and the soft function. The former describes collinear radiation near the hadron, and is defined via matrix elements of non-local operators containing a staple-shaped Wilson line. The latter encodes the soft-gluon effects between the partons. The soft function contains little physical information on incoming hadrons, but has a practical importance: it cancels the rapidity renormalization scale in the beam function. A combination of the two components is necessary to extract the TMDs. It should be noted that both the beam function and the soft function are calculable in lattice QCD.

There have been extensive discussions within the phenomenological community on the factorization related to TMD PDFs and their definitions (see, e.g., Ref.~\cite{Collins:2017oxh}). Recently, such discussions became relevant within the lattice QCD community, due to the increased interest to generalize methods initially developed for PDFs, to access TMD PDFs. Lattice calculations of matrix elements with non-local operators related to TMDs, their renormalization, and the soft function, have recently emerged. However, the field is still at its infancy, and a lot of progress is required. In this article, we briefly summarize the latest progress, most of which is on the theoretical side. More details on the first explorations of TMDs from lattice QCD~\cite{Hagler:2009mb,Musch:2010ka,Musch:2011er,Engelhardt:2015xja,Yoon:2017qzo,Engelhardt:2017miy,Engelhardt:2018zma} can be found in Ref.~\cite{Lin:2020rut}. The most recent studies that are discussed here, can also be found in Refs.~\cite{Ji:2020ect,Lin:2020rut}.

\subsection{Quasi-TMDs approach}

The light-cone TMD PDFs involve matrix elements of non-local operators containing a staple-shaped Wilson line. The latter are infinite and their edges are on, or near the light-cone. Therefore, they cannot be accessed directly in lattice QCD. A way around this issue, is to calculate correlation functions with space-like separated partons, as done for PDFs and GPDs, and then properly match them to their light-cone counterparts. The quasi-distributions approach is the first to be explored for TMDs, and recent work can be found in Refs.~\cite{Ji:2014hxa,Ji:2018hvs,Ebert:2018gzl,Ebert:2019okf,Ebert:2019tvc,Ji:2019sxk,Ji:2019ewn,Vladimirov:2020ofp,Ji:2020ect,Ebert:2020gxr}. 

A prescription to obtain the Collins-Soper evolution kernel non-perturbatively from ratios of quasi-TMDs formed at different momenta, is given in Ref.~\cite{Ebert:2018gzl}. Ref.~\cite{Ebert:2019okf} presents a study of quasi-TMD PDFs and ratios of impact-parameter quasi-TMDs, which can be matched to their light-cone counterparts. In addition, the appropriate matching formula is discussed. Ref.~\cite{Ebert:2019tvc} presents a renormalization prescription and matching of quasi-TMDs as defined in Ref.~\cite{Ebert:2019okf}. A matching kernel of ratios of TMDs for different spin structures is proposed in Ref.~\cite{Ebert:2020gxr}.

\begin{figure}[h!]
\begin{center}
\resizebox{0.47\textwidth}{!}{\includegraphics{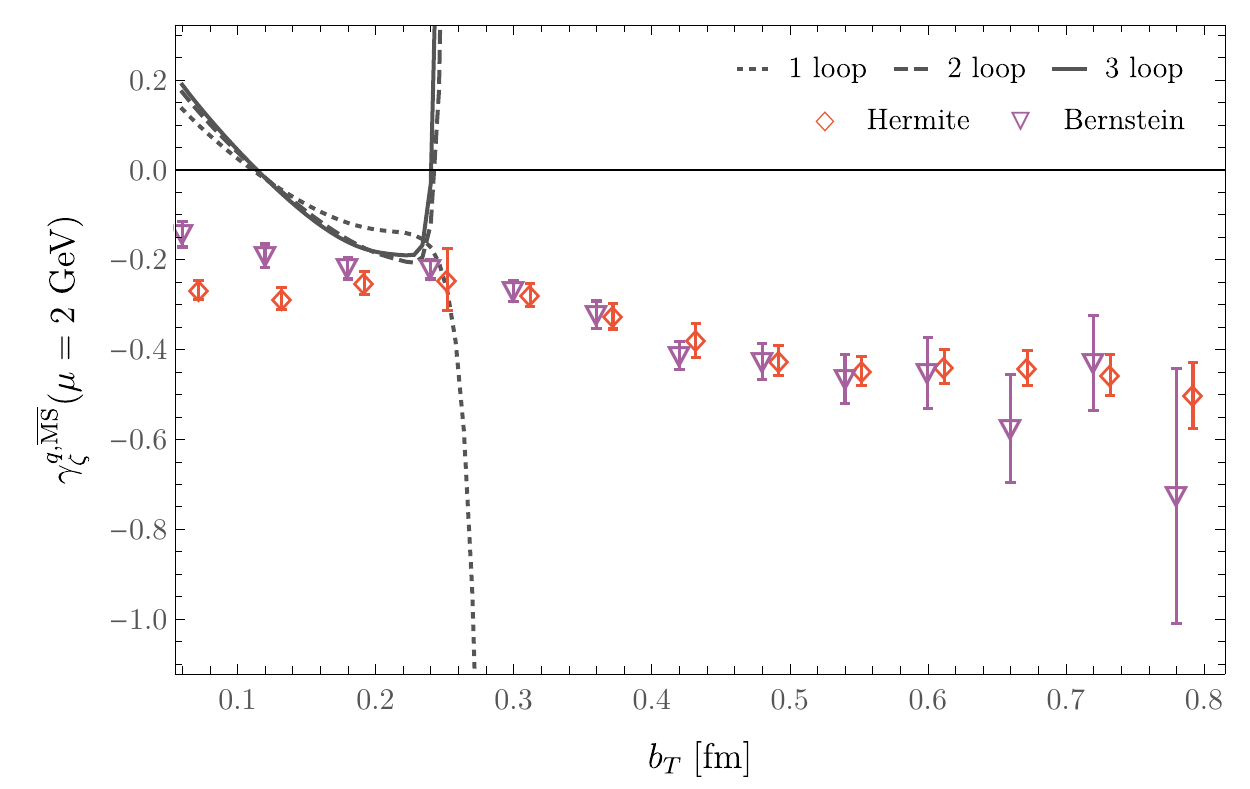}}
\caption{Lattice results on the Collins-Soper evolution kernel as a function of $b_T$. The interpolation of the unsubtracted quasi-TMD PDF using Hermite and Bernstein polynomial bases are shown with red diamonds and purple triangles, respectively. Results from perturbation theory~\cite{Li:2016ctv,Vladimirov:2016dll} are shown with dashed and solid lines. Source: Ref.~\cite{Shanahan:2020zxr}. Article published under the terms of the Creative Commons Attribution 4.0 International license.} 
\label{fig:CS}
\end{center}
\end{figure}

The methodology proposed in Refs.~\cite{Ebert:2018gzl,Ebert:2019okf,Ebert:2019tvc} was implemented numerically in Ref.~\cite{Shanahan:2019zcq} for the renormalization using the gradient flow method, and in Ref.~\cite{Shanahan:2020zxr} for the Collins-Soper kernel. The calculation is performed in quenched LQCD with pion mass of 1.207 GeV. The hadron is boosted with $P_3=1.29,\,1.94,\,2.58$ GeV and for transverse parton momentum $q_T$ in the range of 250 MeV and 2 GeV. The staple extent, $\eta$, is chosen 0.6, 0.7 and 0.8 fm, while the asymmetry in the staple, $b_T$, is between $-(\eta - a)$ and $(\eta - a)$, with $a=0.06$ fm. The extracted Collins-Soper kernel is shown in Fig.~\ref{fig:CS}, together with results from perturbation theory. An agreement is observed only around $b_T\sim 0.2 - 0.3$ fm. This indicades non-negligible systematic effects in the lattice data, such as power corrections in the small-$b_T$ region. The challenges of the inverse problem arise in these calculations too, as the Fourier transform over $b_T$ is required. A way to perform the Fourier transform with limited data is to apply fits on the lattice data, as performed in Ref.~\cite{Shanahan:2020zxr}. However, an extensive study of the systematic uncertainties is necessary to control unwanted effects.

\begin{figure}[h!]
\begin{center}
\resizebox{0.47\textwidth}{!}{\includegraphics{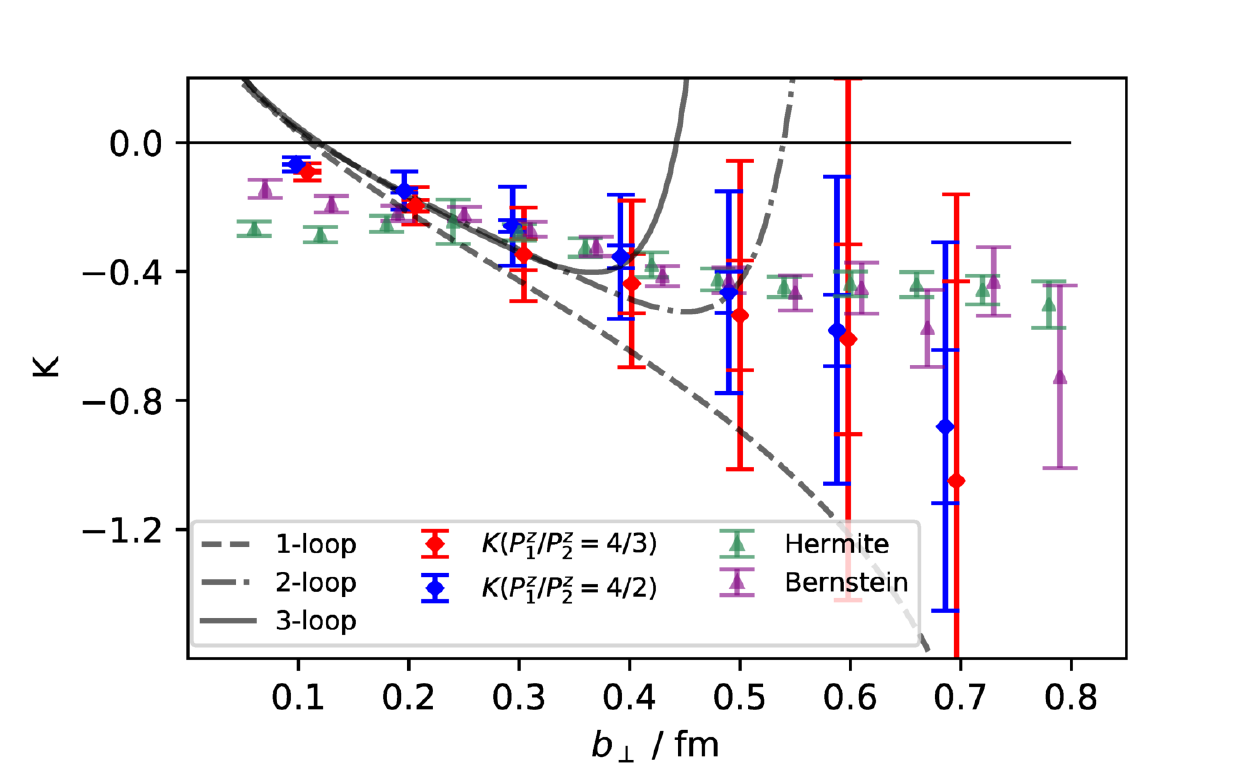}}
\caption{The Collins-Soper kernel from the lattice calculation is shown with blue and red points, and is compared to the quenched calculation of~\cite{Shanahan:2020zxr}. Results from perturbation theory~\cite{Li:2016ctv,Vladimirov:2016dll}  are shown with solid and dashed lines, for 1-loop, 2-loop and 3-loop estimates.
Source: Ref.~\cite{Zhang:2020dbb}. Reprinted based on the arXiv distribution license.} 
\label{fig:CS_kernel}
\end{center}
\end{figure}

The study of the soft function as defined in the Collins-Soper TMD factorization is discussed in Ref.~\cite{Ji:2019sxk}. One of the findings is that the soft function can be defined as the form factor of a pair of color sources traveling at almost the speed of lights, and can be calculated using lattice Heavy-Quark Effective Theory (HQET). However, this is of little practical use for lattice calculations. An alternative proposal is to extract the soft function through the factorization of a fast-moving light-meson form factor, which can be combined with the quasi-TMD wavefunction. In the follow-up calculation of Ref.~\cite{Ji:2019ewn}, the matching kernel is extracted, based on the definition of Ref.~\cite{Ji:2019sxk}. Ref.~\cite{Zhang:2020dbb} presents a lattice calculation of the rapidity-independent part of soft function presented in the transverse coordinate space, indicated as $b_\perp$. The calculation is performed using one $N_f=2+1$ CLS ensemble of clover fermions with pion mass 333 MeV in the sea sector and 547 MeV in the valence sector. The momentum boost is up to $2.1$ GeV. The final results for the two ensembles are shown in Fig.~\ref{fig:CS_kernel}, and are compared with perturbative results, as well as the results from Ref.~\cite{Shanahan:2020zxr}. We note that the comparison is qualitative, as the latter calculation is quenched, and at a very heavy pion mass. This has the benefits of small statistical uncertainties. An agreement is observed between the various lattice results, within the errors. The results of  Ref.~\cite{Zhang:2020dbb} are compatible with the 3-loop perturbative results up to $b_T\sim 0.4$ fm.

\vspace*{0.5cm}
\section{Closing Remarks}
\label{sec:Discussion}

The work presented in this review is a demonstration of the advancement of lattice QCD in novel methods of obtaining PDFs, GPDs and TMD PDFs that, only a decade ago, were considered impossible to calculate ``directly''. In particular, we have demonstrated that calculations of PDFs have reached an advanced level, with dedicated studies of various sources of systematic uncertainties. Given the progress, it is possible that precision calculations with controlled uncertainties will be available in the next few years. Gluon PDFs and flavor-singlet quark PDFs are actively being pursued with very promising results. This direction progresses more slowly than the non-singlet PDFs, as the calculation of disconnected contributions of non-local operators and boosted hadrons are computationally very challenging. The natural extension of the methods to access $x$-dependent distribution functions, is studies of GPDs. Interestingly, only exploratory studies are available, providing the $x$-dependence of GPDs for selected values of the momentum transfer at zero skewness, or limited values of nonzero skewness. This is due to the fact that the introduction of momentum transfer comes with several complications, most of them computational. The above make the computation of GPDs much more challenging than the form factors, due to the frame-dependence of the former. Serious effort is invested in studies of TMD PDFs, with most of the work being on theoretical developments. The lattice calculations are still very preliminary, and some of them at very heavy pion mass. A lot more numerical explorations are expected to follow. 

As concluded from the main text, the various methods to access $x$-dependent distribution functions are based on different theoretical formulations and are susceptible to different systematic uncertainties. Regardless, all methods pursue the same physical quantities, even though the factorization procedure is realized differently. Some methods rely directly on large hadron momentum, $P_3$, and others on large Ioffe time, $z P_3$. Regardless on the classification of the factorization (LaMET, or short-distance), the calculations of matrix elements with boosted hadrons and non-local operators are computationally intensive. It is apparent that there are limitations to the value of $P_3$ due to the computational cost and the value of $a$. For coordinate space factorization, there are also limitations on $z$, which has to be small and within the perturbative region. Since there is a need for large $P_3$, or large $z P_3$, it is undoubtedly desirable to explore what the current capabilities of the lattice calculations are. Often, there is the misconception that large momentum can be achieved with controlled uncertainties. Instead of making assumptions based on phenomenological models or otherwise, we take a close look at the current calculations to address the question whether it is feasible to reach high momentum without compromising the reliability of the results. Such a question becomes more pressing for simulations near or at the physical point. 

\begin{table}[h!]
\caption{Parameters of various lattice calculations and comparison of the noise-to-signal ratio.}
\label{tab:param}       
\begin{tabular}{lllll}
\hline\noalign{\smallskip}
Ref. & $m_\pi$(MeV) & $P_3$(GeV) & $\displaystyle\frac{n}{s}\Big{|}_{z{=}0}$  \\[0.5ex]
\noalign{\smallskip}\hline\noalign{\smallskip}
quasi/pseudo~\cite{Alexandrou:2019lfo,Bhat:2020ktg} 
&\,\, 130        &\,\, 1.38    &  6$\%$       \\[2.5ex]
pseudo~\cite{Joo:2020spy}   
&\,\, 172       &\,\, 2.10      &   $8\%$    \\[2.5ex]
current-current ~\cite{Sufian:2020vzb}   
&\,\, 278      &\,\, 1.65     &  $19\%\,^{\bf{\star}}$   \\[2.5ex]
quasi~\cite{Izubuchi:2019lyk}
&\,\, 300        &\,\, 1.72     &  $6\%\,^{\bf{\dagger}}$       \\[2.5ex]
quasi/pseudo~\cite{Gao:2020ito}   
&\,\, 300      &\,\, 2.45   &   $8\%\,^{\bf{\dagger}}$   \\[2.5ex]
quasi/pseudo~\cite{Fan:2020nzz}   
&\,\, 310      &\,\,  1.84    &   $3\%\,^{\bf{\dagger}}$   \\[1ex]
------ \\
twist-3~\cite{Bhattacharya:2020cen}     
&\,\, 260       &\,\, 1.67     &  $15\%$      \\[1ex]
------\\
$s$-quark quasi~\cite{Alexandrou:2020uyt}    
&\,\, 260       &\,\, 1.24     &   $31\%$   \\[2.5ex]
$s$-quark quasi~\cite{Zhang:2020dkn}
&\,\, 310       &\,\, 1.30     & $43\%\,^{\bf{\star\star}}$    \\[2.5ex]
gluon pseudo~\cite{Fan:2020cpa}     
&\,\, 310       &\,\, 1.73     & $39\%$        \\[1ex]
------ \\
quasi-GPDs~\cite{Alexandrou:2020zbe}\\$-t{=}0.69$GeV$^2$     
&\,\, 260       &\,\,  1.67     &    $23\%$     \\[2.5ex]
quasi-GPDs~\cite{Chen:2019lcm}\\$-t{=}0.92$GeV$^2$     
&\,\, 310       &\,\,  1.74     &    $59\%$   \\
\noalign{\smallskip}\hline 
\end{tabular}
\vspace*{0.25cm}

{\footnotesize{$\bf{\dagger}$ At $T_{\rm sink}<1$ fm. \\[0.5ex]
${\bf{\star}}$ At smallest $z$ value used, $z=2$. \\[0.5ex]
{\bf{$\star\star$}} At maximum value of imaginary part, $z=4$.}}
\end{table}
Table~\ref{tab:param} presents some of the parameters of recent calculations presented in this article, either for the pion or the nucleon. The main focus is on the noise-to-signal ratio, $\displaystyle\frac{n}{s}$, at the maximum value of the matrix elements. For the real part of the matrix elements the peak is at $z=0$, while for the imaginary part is at some intermediate value of $z$ depending on the quantity under study. As most calculations include multiple values of $P_3$, we present $\displaystyle\frac{n}{s}$ at the highest $P_3$, which will provide some indication of the challenges to increase $P_3$ for non-local operators. It should be emphasized that the noise-to-signal percentages should not be directly compared between methods, as the focus of each approach is different. For example, the pseudo-PDF method and current-current correlators method utilize all values of $P_3$, while for the quasi-PDFs analysis, one uses the data at the highest possible momentum. Therefore, the control of statistical uncertainties in the former methods can and is achieved at low momenta. For example, if one would analyze the data of Ref.~\cite{Joo:2020spy} within the quasi-PDFs method, the maximum momentum which leads to reliable renormalized results for a wide range of $z$ values is $P_3\sim1.3$ GeV for the 172 MeV ensemble. However, the results at $P_3=2.1$ GeV are accurate enough for small values of $z$ within the pseudo-ITDs renormalization scheme. As can be seen, the picture is consistent from all calculations: $P_3$ close or beyond 2 GeV cannot be reached reliably. This is particularly true for ensembles with a pion mass near or at the physical point. 

The general picture of Table~\ref{tab:param} should not be viewed as a major disadvantage of the field, but as a motivation to develop and explore novel approaches, innovative techniques and algorithms~\footnote{During a panel discussion at the Lattice Symposium in 2001 that took place in Berlin, the infamous ``Berlin Wall plot''~\cite{Bernard:2002pd} was shown, concluding that simulations at the physical point were unreachable. However, this changed around 2008 with the development of new algorithms and novel computer architecture~\cite{Jansen:2008vs}.}. Such an example is the momentum smearing method~\cite{Bali:2016lva}, first explored in Ref.~\cite{Alexandrou:2016jqi} for non-local operators. Since then, it has been integrated into all calculations. This direction has been instrumental in allowing access to higher values of $P_3$. A recent exploration of the momentum-smearing within the distillation framework can be found in Ref.~\cite{Egerer:2020hnc}. We will soon enter the exascale computing era, which promises more computing power. The combination of software and hardware developments will certainly benefit calculations of non-local operators, leading to precision calculations with controlled uncertainties.

The purpose of this review is to highlight the many successes of the field, and point out the aspects that need further exploration. Regardless of the necessary improvements, there is no doubt that accessing $x$-dependent parton distributions from lattice QCD is a major success towards understanding the complex structure of hadrons, and gains visibility from the experimental and theoretical nuclear physics communities. 

\vspace*{1.5cm}
\noindent
\centerline{\textbf{Acknowledgements}}
\vspace*{0.1cm}

I would like to thank the organizers of the 38$^{\rm th}$ International Symposium on Lattice Field Theory (Lattice 2020) for the invitation to write this review, in lieu of a plenary talk~\footnote{The Conference was cancelled due to the COVID-19 pandemic.}. The article will be included in the EPJA special issue ``Lattice Field Theory during the COVID-19 pandemic''.

This review is dedicated to the junior researchers involved in the calculations presented here for their outstanding work.
I would like to thank K. Cichy, X. Ji, J. Karpie, A. Metz, A. Radyushkin,  R. Suffian, and Y. Zhao for interesting discussions. I am particularly grateful to K. Cichy for providing feedback that improved the manuscript.

The Author acknowledges financial support by the U.S. Department of Energy, Office of Nuclear Physics, Early Career Award under Grant No.\ DE-SC0020405. 

\vspace*{.5cm}
 \bibliographystyle{epj}
 \bibliography{references}

\end{document}